\documentclass[amssymb,amsmath,eqsecnum,showpacs,nofootinbib,twocolumn] {revtex4}
\usepackage{color}
\usepackage{footnote}
\usepackage{graphicx}
\usepackage{ulem}
\usepackage{physics}
\usepackage{verbatim}
\usepackage{bigints}
\usepackage{float} 
\usepackage[colorlinks]{hyperref}

\hypersetup{
	colorlinks,
	citecolor=blue,
	filecolor=gray,
	linkcolor=blue,
	urlcolor=blue
}

\begin{document}

\author{Jonathan Claros$^{1,2}$ and Emanuel Gallo$^{1,2}$
}

\newcommand{\red}[1]{\textcolor{red}{#1}}
\newcommand{\blue}[1]{\textcolor{blue}{#1}}

\affiliation{$^1$ Universidad Nacional de Córdoba, Facultad de Matemática, Astronomía, Física y Computación, Grupo de Relatividad y Gravitación; Córdoba, Argentina.\\ $^2$ Consejo Nacional de Investigaciones Científicas y Técnicas,\\ CONICET, IFEG. Córdoba, Argentina. }

\title{
Fast and accurate analytical formulas for light propagation in general static, spherically symmetric spacetimes}

\begin{abstract}

In this article, we extend our previously presented analytical formulas \cite{Claros:2024atw} for describing light rays passing near or emitted in the vicinity of compact objects to a broader class of spherically symmetric, static spacetimes, including the Johansen–Psaltis and Rezzolla-Zhidenko metric families. The generalized formulas retain the simplicity and accuracy of the original approach while allowing for more general deviations from Schwarzschild geometry. These expressions provide an approximate yet accurate mapping between emission points and the image plane of an asymptotic observer, enabling fast analytical computations of accretion disk images, polarization of the emitted radiation, luminosity curves associated with pulsars, and other related applications. As examples, we compute isoradial curves for several metric families and the Stokes parameters $Q$ and $U$ for a hot spot orbiting near a black hole described by one of the studied metrics, presenting the corresponding polarization ($QU$) curves.

\end{abstract}

\maketitle
\section{Introduction} 

In recent years, the study of compact objects, such as black holes and neutron stars, has become increasingly relevant in astrophysics. The Event Horizon Telescope (EHT) collaboration provided the first direct images of the black holes M87* and Sgr A* \cite{EventHorizonTelescope:2022wkp,EHT:2019:1}. Concurrently, the polarization analysis of ``hotspots" orbiting black holes has allowed for the characterization of the composition and dynamics of their accretion disks \cite{2005MNRAS.363..353B,refId0,GRAVITY:2020hwn,GRAVITY:2023avo}. Regarding neutron stars, the observation of pulse profiles from radio and/or X-ray emissions, along with polarization studies \cite{Gnarini:2022geo,Bobrikova:2024soh,Benacek:2025wgp} using instruments like NuSTAR \cite{NuSTAR:2013yza}, NICER, and IXPE, has enabled the estimation of their mass-radius relations. This imposes crucial constraints on their internal composition \cite{Bogdanov:2015tua} and allows for the analysis of the effects of the surrounding plasma environment on light \cite{Briozzo_2022,Kobialko:2025sls}. Therefore, studying light in the vicinity of these compact objects has become a key tool for testing General Relativity and alternative theories of gravitation in the strong-field regime \cite{Pantig:2022ely,Shahzadi:2022rzq,Freire:2024adf}.

In studies of light propagation near compact objects, the standard tool is the ray-tracing of null geodesics, which enables the investigation of observable photon trajectories. However, this method is computationally expensive due to the numerical integration involved. For this reason, several approximate analytical approaches have been developed to perform this task more efficiently~\cite{Beloborodov_2002,Semerak:2014kra,LaPlaca:2019rjz,Poutanen:2019tcd,Gao:2025arj}. Analytical approximations of this kind have proven remarkably effective in modeling a wide range of phenomena involving neutron stars and black holes~\cite{Poutanen:2003yd,Viironen:2004ze,Gierlinski:2004tu,Kulkarni:2005cs,Suleimanov:2006pk,vanAdelsberg:2006uu,Poutanen:2006hw,Bogdanov:2006zd,Morsink:2007tv,Ho:2008bq,Ibragimov:2009js,Dipanjan:2011ef,Poutanen:2013xaa,Potekhin:2014fla,Mohan:2015jsa,Watts:2016uzu,Perego:2017fho,Mushtukov:2017ubg,Ascenzi:2024osa,Saathoff:2024pzk,Cardenas-Avendano:2022csp,Loktev:2023cty,Markozov:2023zxq,Ahlberg:2023iza,EventHorizonTelescope:2021btj,Hu:2022ehk,Loktev:2020tin,BG_2023,Junior:2024lvj}.

Among these approaches, the Beloborodov formula (Eq.~(1) in~\cite{Beloborodov_2002}) stands out. In the Schwarzschild spacetime, it provides a relation between the emission angle of a light ray; measured with respect to the radial direction; and the angular (and radial) position at which the same point is observed in the asymptotic region. Despite its success, this formula has some intrinsic limitations: it loses accuracy for emission angles exceeding $90^{\circ}$, it assumes exact spherical symmetry (thus neglecting spin-induced spacetime deformations), and it is restricted to the Schwarzschild metric.

Some of these shortcomings have been mitigated in later works. In particular, Poutanen~\cite{Poutanen:2019tcd} proposed a refined expression; still within the Schwarzschild geometry; that significantly improves the treatment of large-angle emission and has been successfully applied in various astrophysical contexts. 

{ Moreover, comparative studies between approximate analytical treatments of photon propagation in a non-spinning Schwarzschild spacetime and full Kerr ray-tracing simulations indicate that, for moderate black-hole spins, both the resulting direct images and their associated polarization signatures remain in close quantitative agreement~\cite{Cardenas-Avendano:2022csp,Loktev:2023cty}.} Notably, the Poutanen approximation has been implemented in the ARTPOL algorithm, which enables fast analytical tracing of polarized light rays with accuracy comparable to numerical ray-tracing for Kerr black holes with spin parameters up to $a_* \simeq 0.94$, while being over four orders of magnitude faster~\cite{Loktev:2023cty}.

Nevertheless, these analytical formulas remain restricted to the Schwarzschild spacetime and therefore cannot be employed to test, through the associated observables, the strong-field regime predicted by alternative theories of gravity. To address this limitation, in our previous work~\cite{Claros:2024atw} we proposed a method to derive more general analytical expressions valid for spherically symmetric spacetimes, thereby extending the approaches of Beloborodov and Poutanen. In that work, we showed the accuracy and versatility of the proposed analytical formulas in modeling a wide class of black hole spacetimes, with applications to the generation of accretion-disk images and to analytical studies of polarimetric signatures. However, the applicability of these improved expressions remained limited to spherically symmetric metrics that satisfy the condition $g_{tt} = -1/g_{rr}$ in Schwarzschild-like coordinates. Even though a wide class of metrics arising in alternative theories of gravity satisfy this relation, it is not universal. Indeed, even assuming the Einstein equations hold, its validity still requires the energy--momentum tensor to satisfy specific conditions~\cite{Salgado:2003ub, Gallo:2003rt}.

{ The primary goal of this paper is to overcome this limitation by extending and generalizing our previous analytical framework to describe photon trajectories in generic static and spherically symmetric spacetimes. The resulting expressions provide a consistent treatment of direct rays; namely, those completing less than half an orbit around the compact object, for which the emission angle $\psi$ of a emission point $p$ as shown in Fig.~\ref{fig:lightray} remains smaller than $\pi$. A more detailed discussion of this geometric configuration will be presented in the next section. Consequently, across a broad class of gravitational backgrounds, the new formulas significantly enhance the applicability of semi-analytical modeling in the strong-field regime.}

The organization of this paper is as follows. Section~\ref{sec:approximations} presents the analytical approximations~\eqref{eq:our_aprox_full} and \eqref{eq:our_aprox_belo} together with their derivation. In Sec.~\ref{Sec:Testing}, we test the accuracy of these approximations. Section~\ref{sec:metrics_fam} applies them to parameterized metrics such as the nonrotating Johannsen--Psaltis~\cite{PhysRevD.83.124015} and the Rezzolla--Zhidenko~\cite{Rezzolla:2014mua} metrics. Section~\ref{sec:metrics_part} considers a parameterization of the Einstein--Maxwell--dilaton--axion metric~\cite{PhysRevLett.74.1276}. In Sec.~\ref{sec:applications}, we present two applications of the approximations: Subsec.~\ref{sec:isoradial_curves} generates equatorial isoradial curves for the above metrics, while Subsec.~\ref{sec:qu_diagrams} computes the $QU$ loop diagrams produced by synchrotron-emitting hotspots in the Einstein-Maxwell-dilaton-axion spacetime. We conclude with some final remarks in Sec.~\ref{sec:final_remarks}. Throughout this paper we adopt the $(-,+,+,+)$ signature and geometrized units with $G = c = 1$.

\section{General approximations} \label{sec:approximations}

\subsection{Preliminaries}
\begin{figure} [h!]
    \centering
    \includegraphics[scale=0.6]{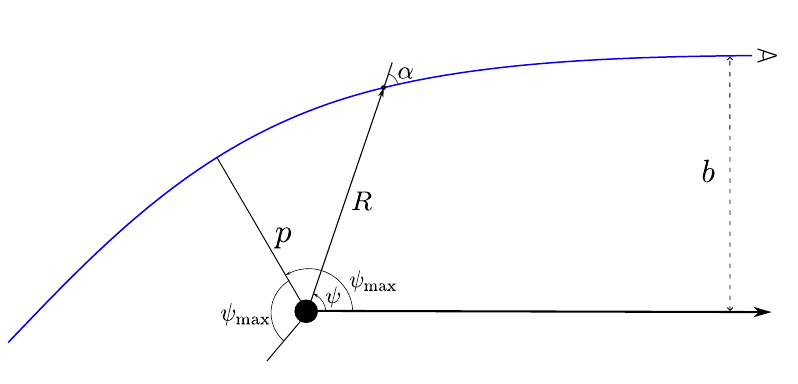}
    \caption{{Figure reproduced from Ref.~\cite{Claros:2024atw}, Fig.~1. It illustrates} a light ray traveling from the vicinity of a compact object to a distant observer. The ray reaches the observer, located in the asymptotic region ($\psi = 0$), with an impact parameter $b$. The ray's point of closest approach (periastron) occurs at the radial coordinate $r = p$. At a given coordinate $r=R$, $\alpha$ represents the angle between the radial direction and the light ray’s path.}
    \label{fig:lightray}
\end{figure}

Let us consider a light ray passing near a compact object. This ray could be emitted by an accretion disk surrounding the object, by a star, by an orbiting hotspot, by an polar cap on neutron star surface among other possibilities. At a given point $(R, \psi)$ of this trajectory, the ray forms an angle $\alpha$ with respect to the radial direction. If there is a light emitter at $(R, \psi)$, this angle is referred to as the emission angle. The ray eventually reaches an asymptotic observer with impact parameter $b$, as illustrated in Figure \ref{fig:lightray}.

By defining $x = 1 - \cos \alpha$ and $y = 1 - \cos \psi$, an approximate yet accurate relation between $\alpha$ and the radial coordinates $(R, \psi)$ was proposed by Beloborodov \cite{Beloborodov_2002} for the Schwarzschild spacetime:

\begin{equation} \label{eq:belo_aprox}
    x=(1-u)y, \qquad u:=\frac{r_{\text{sch}}}{R},
\end{equation}
where $r_{\text{sch}}$ denotes the Schwarzschild radius. Poutanen and Beloborodov \cite{Poutanen:2006hw} later extended this approximation by including corrections of order $y^3$, and subsequent work \cite{Poutanen:2019tcd} introduced an empirical fitting term proportional to $y^2\left[\ln(1 - y/2) + y/2\right]$ to further improve the original accuracy of Eq.~\eqref{eq:belo_aprox}. The resulting approximation reads:

\begin{equation} \label{eq:Poutanen}
    x=(1-u)y\left\{1+\frac{u^2}{112}y^2-\frac{e}{100}uy\left[\ln(1-\frac{y}{2})+\frac{y}{2}\right]\right\}.
\end{equation}

In our previous work \cite{Claros:2024atw}, generalizations of Eqs.~\eqref{eq:belo_aprox} and \eqref{eq:Poutanen} were derived by considering a more general spherically symmetric spacetime of the form 
\begin{equation} \label{eq:metric_spheric1}
    ds^2=-A(r)dt^2+\frac{dr^2}{A(r)} + r^2(d\theta^2+\sin^2\theta d\phi^2).
\end{equation}
The resulting formulas were expressed in terms of integrals of functions depending on the compactness parameter $u$. 

\subsection{Approximation formulas for more general metrics}

Now, let us consider a static and spherically symmetric spacetime described by the metric
\begin{equation} \label{eq:metric_spheric}
    ds^2=-A(r)dt^2+B(r)dr^2 + r^2(d\theta^2+\sin^2\theta d\phi^2);
\end{equation}
{which differs from Eq.~\eqref{eq:metric_spheric1} by allowing for a more general function $B(r)$, instead of imposing $B(r) = A^{-1}(r)$.}
Because the orbits of interest are confined to a plane, we can, without loss of generality, perform an angular coordinate transformation $(\theta, \phi) \rightarrow (\tilde{\theta}, \psi)$ such that the light trajectory lies in the equatorial plane, defined by $\tilde{\theta} = \pi/2$. In this coordinate system, the metric takes the form
\begin{equation} \label{eq:metric_light}
ds^2 = -A(r)dt^2 + B(r)dr^2 + r^2(d\tilde{\theta}^2 + \sin^2\tilde{\theta}d\psi^2).
\end{equation}
 
As shown in Fig.~\ref{fig:lightray}, once the light ray passes the periastron $p$-where the emission angle satisfies $\alpha = \pi/2$-and moves toward the asymptotic region, we have $\dot{\psi} < 0$ and $\dot{r} > 0$. In this region, the null geodesic equation can be written as
\begin{equation} \label{eq:dpsi_dr}
    \frac{d\psi}{dr} = -\frac{1}{r^{2}}\left[\frac{1}{A(r)B(r)}  \left(\frac{1}{b^{2}} - \frac{A(r)}{r^2}\right)\right]^{-1/2},
\end{equation}
whose integration yields

\begin{equation} \label{eq:psi_def}
    \psi(R,b)=\int_{R}^{\infty}\frac{dr}{r^{2}}\left[\frac{1}{A(r)B(r)}  \left(\frac{1}{b^{2}} - \frac{A(r)}{r^2}\right)\right]^{-1/2}.
\end{equation}
After substituting $r = R/u'$ and expressing the impact parameter as~\cite{Pechenick_1983,Beloborodov_2002},
\begin{equation} \label{eq:impact_parameter}
    b=\frac{R}{\sqrt{A(R)}}\sin\alpha,
\end{equation}
 together with the replacement $\sin\alpha = \sqrt{1 - \cos^2\alpha}$, Eq.~\eqref{eq:psi_def} becomes
\begin{equation} \label{eq:belo-ap1}
    \psi(R,\alpha)=\bigintss_{0}^{1}\frac{\sqrt{A(\frac{R}{u'})B(\frac{R}{u'})}\sqrt{1-\cos^2\alpha}}{\sqrt{A(R)-A(\frac{R}{u'})u'^2(1-\cos^2\alpha})}du'.
\end{equation}

Equation~\eqref{eq:belo-ap1} is valid only for $\alpha \in \left[0,\tfrac{\pi}{2}\right]$. For  $\alpha>\tfrac{\pi}{2}$, it is calculated as~\cite{Poutanen:2019tcd}:

\begin{equation} \label{eq:psi_p}
    \psi(R,\alpha)=2\psi_{\text{max}}-\psi(R,\pi-\alpha),
\end{equation}
where second term uses~\eqref{eq:belo-ap1} and $\psi_{\text{max}}=\psi(p,\tfrac{\pi}{2})$ with $p$ the periastron which is obtained from $p^2-b^2A(p)=0$.

To derive our analytical approximations, we define $v^2 = 1 - \cos\alpha$, and expand the integrand of Eq.~\eqref{eq:belo-ap1} around $v = 0$:

\begin{widetext}
\begin{equation} \label{eq:psi_expand}
    \psi(R,v)\approx\frac{\sqrt{2}}{\sqrt{A(R)}} \left[ I_1 v -\frac{I_2}{4A(R)}v^3 - \frac{I_3}{32A(R)^2}v^5 \right]+\mathcal{O}(v^{7}), \\
\end{equation}
with
\begin{subequations} \label{eq:integrals_i}
    \begin{align}
    &I_1:=\mathcal{I}_1, \\
    &I_2:=-4\mathcal{I}_2+A(R)\mathcal{I}_1,\\
    &I_3:=-48\mathcal{I}_3+24A(R)\mathcal{I}_2+A(R)^2\mathcal{I}_1;
    \end{align}
\end{subequations}
where 
\begin{eqnarray} \label{eq:int-main}
    \mathcal{I}_n&=&\int^1_0du'\sqrt{A(u')B(u')}[A(u')u'^2]^{n-1},
\end{eqnarray}

\end{widetext}
and the functions of the metric $A(u'),B(u')$ are obtained from $A(r),B(r)$ after the substitution $r=R/u'$.

A notable feature of the integrals $\mathcal{I}_1$, $\mathcal{I}_2$, and $\mathcal{I}_3$ is that they can be evaluated analytically for a nontrivial class of metrics, thereby yielding an approximate analytical expression for $\psi$, in contrast to the original integral~\eqref{eq:belo-ap1}.

Taking Eq.~\eqref{eq:psi_expand} into account, we perform a Taylor expansion of $y = 1 - \cos\psi$ in powers of $v$. This expansion contains only even powers of $v$. Using that $x = v^2$, we obtain

\begin{equation} \label{eq:y_expand}
    y=A_1 x+A_2x^2+A_3x^3+\mathcal{O}(x^4),
\end{equation}
with coefficients 
\begin{eqnarray} \label{eq:coeff_a}
A_1&=&\frac{I_1^2}{A(R)}, \nonumber \\
A_2&=&-\frac{I_1\left(I_1^3+3I_2\right)}{6A(R)^2},\\
A_3&=&\frac{8I_1^6+120I_2 I_1^3 -45I_3 I_1 +45I_2^2}{720A(R)^3}. \nonumber
\end{eqnarray}
Since the inverse relation is of interest-namely, expressing $x$ as a power series in $y$-we consider an expansion of the form
\begin{equation} \label{eq:x_expand}    x=B_1y+B_2y^2+B_3y^3+\mathcal{O}(y^4).
\end{equation}

Substituting Eq.~\eqref{eq:x_expand} into Eq.~\eqref{eq:y_expand} and equating coefficients of like powers of $y$ yields the following relations:

\begin{eqnarray} \label{eq:coeff_b}
B_1&=&\frac{1}{A_1}, \nonumber \\
B_2&=&-\frac{A_2}{A^3_1},\\
B_3&=&\frac{2A^2_2-A_1A_3 }{A^5_1}. \nonumber
\end{eqnarray}

Finally, with these coefficients, the following explicit approximation is obtained 
\begin{equation} \label{eq:our_aprox_o3}
\begin{split}
    x=&\frac{A(R)}{I_1^2}y+\frac{A(R)(I_1^3+3I_2)}{6I_1^5}y^2\\
    &+\frac{A(R)(32I_1^6+120I_1^3I_2+45I_3I_1+315I_2^2)}{720I_1^8}y^3.
    \end{split}
\end{equation}

It is worth noting that this expression has been derived fully analytically and, as discussed in the next section, provides an excellent approximation as long as $\psi$ is not too close to $\pi$. However, in analogy with Eq.~\eqref{eq:Poutanen}, its accuracy can be further improved by adding an empirical correction term of the form
\begin{equation}
-\frac{e}{100}\gamma(R)y^{2}\left[\ln\left(1 - \frac{y}{2}\right) + \frac{y}{2}\right],
\end{equation}
to Eq.~\eqref{eq:our_aprox_o3}, where $\gamma(R)$ is a function of $R$ whose explicit form depends on the metric under consideration. For the specific metric considered in Sec.~\ref{sec:metrics_part} (the Einstein--Maxwell--dilaton--axion metric), we adopt
\begin{equation}\label{eq:gammaviejo}
\gamma(R) = A(R)(1 - A(R)),
\end{equation}
whereas for metric families, we consider two different expressions for $\gamma(R)$. In particular, for the Johannsen–Psaltis metric~\cite{PhysRevD.83.124015}, we set
\begin{equation}\label{eq:gamma_johan}
    \gamma(R) = \left(\sqrt{\tfrac{A(R)}{B(R)}}\right)\left(1 - \frac{1}{\sqrt{B(R)}}\right)\left(1 + \sqrt{A(R)}\right),
\end{equation}
while for the Rezzolla-Zhidenko family \cite{Rezzolla:2014mua} we use:

\begin{equation}\label{eq:gamma_rezzo_zhi}
\gamma(R) = A(R)\left(1 - \frac{1}{B(R)}\right).
\end{equation}

Any of these expressions for $\gamma(R)$ reproduce the empirical term used in Ref.~\cite{Claros:2024atw} when $B(R) = 1/A(R)$ is assumed. Moreover, they reduce to the one proposed by Poutanen~\cite{Poutanen:2019tcd} in the case of the Schwarzschild metric.

Thus, the final expression relating the emission angle $\alpha$ to $(R, \psi)$ is given by
\begin{widetext}
\begin{equation} \label{eq:our_aprox_full}
    x=\frac{A(R)y}{I_1^2}+\frac{A(R)(I_1^3+3I_2)y^2}{6I_1^5}+\frac{A(R)(32I_1^6+120I_1^3I_2+45I_3I_1+315I_2^2)y^3}{720I_1^8}-\frac{e}{100}\gamma(R)y^2\left[\ln\left(1 - \frac{y}{2}\right) + \frac{y}{2}\right].
\end{equation}
\end{widetext}

This new formula includes, as particular cases, the Poutanen expression~\cite{Poutanen:2019tcd} and our earlier results in Ref.~\cite{Claros:2024atw}.

In the linear approximation, we have
\begin{equation} \label{eq:our_aprox_belo}
    x = \frac{A(R)}{\left(\int^1_0 du' \sqrt{A(u') B(u')}\right)^2} y,
\end{equation}
which for the particular case of a Schwarzschild metric reduces to the well-known Beloborodov formula Eq.~\eqref{eq:belo_aprox}. For this reason, in this work Eq.~\eqref{eq:our_aprox_belo} will be referred to as the generalized Beloborodov formula. As will be shown below, this compact expression can also be successfully applied to describe light rays in a nontrivial class of astrophysical configurations as observed from infinity.
{ What is most interesting and surprising about all these analytical approximations (including the Beloborodov and Poutanen formulas) is that they provide an approximate relation between $\alpha$ and $\psi$ that remains valid regardless of whether the orbit passes through periastron. In numerical integrations, by contrast, one must first locate the periastron and determine the value $\psi_{\text{max}}$, so that the integral in Eq.\eqref{eq:belo-ap1} is used for $\psi \leq \psi_{\text{max}}$ and the expression in Eq.\eqref{eq:psi_p} is used for $\psi > \psi_{\text{max}}$. Remarkably, in the analytical approximations the relation between $\psi$ and $\alpha$ holds throughout the entire trajectory; as long as $\psi \lesssim \pi$; without the need to compute either the periaxis $p$ or $\psi_{\text{max}}$. For this reason, as we will see in the comparisons between numerical and analytical results in the next section, the numerical integration must be performed in two segments, whereas the analytical expression is single and given by Eqs.\eqref{eq:our_aprox_full} or \eqref{eq:our_aprox_belo}.

}

It should be noted that, although analytical expressions for the integrals $\mathcal{I}_1$, $\mathcal{I}_2$, and $\mathcal{I}_3$ are derived in this work, they can still be evaluated numerically for a given set of spacetime parameters and a specific value of $R$ in cases where an analytical treatment is not feasible. In contrast, the exact expressions given by Eqs.\eqref{eq:belo-ap1} and \eqref{eq:psi_p} must be solved numerically not only for each choice of parameters and $R$, but also for every value of $\alpha$ at a given $R$. Moreover, after generating a corresponding lookup table between $\psi$ and $\alpha$, this relation must be inverted to recover the emission angle as a function of $\psi$. Hence, the proposed approximation may also prove useful in such scenarios. In this paper, however, we restrict our analysis to specific metrics, or families of metrics, for which these integrals can be obtained in fully analytical form.

\section{Validation of the Analytical Approximation} \label{Sec:Testing}

We now test the accuracy of Eqs.~\eqref{eq:our_aprox_full} and~\eqref{eq:our_aprox_belo} by comparing them with the numerical results obtained from Eq.~\eqref{eq:psi_def} for $\alpha < \pi/2$ and from Eq.~\eqref{eq:psi_p} for $\alpha > \pi/2$. For this purpose, we consider two widely used parametric families describing black holes-the Johannsen--Psaltis and Rezzolla--Zhidenko metrics-together with a specific solution, the Einstein--Maxwell--dilaton--axion metric.
 The comparison is performed for a value of $R$ close to the event horizon $r_H$, specifically at $R = 1.1\,r_H$, and for the set of radii $R = \{1.01\,r_c,\, 1.5\,r_c,\, 3\,r_c,\, 4.5\,r_c\}$, which correspond to different locations relative to the photon-sphere radius $r_c$, obtained as the solution of~\cite{Chandra:1983,Cardoso:2008bp}: 
\begin{equation}
    2A(r_c) - r_c \tfrac{dA(r)}{dr}\Big|_{r_c} = 0.
\end{equation}

For all metrics considered, { `{Approx.~T}'} denotes the approximation given by Eq.~\eqref{eq:our_aprox_full}, corresponding to the Taylor expansion supplemented with an empirical term, whereas {`{Approx.~L}'} refers to the linearized approximation derived from the generalized Beloborodov formula, Eq.~\eqref{eq:our_aprox_belo}. The results are compared with the numerical integration of $\psi$, from which the inverse relation $\alpha = \alpha(\psi)$ is obtained numerically, using Eq.~\eqref{eq:psi_def} for $\alpha < \pi/2$ and Eq.~\eqref{eq:psi_p} for $\alpha > \pi/2$.

\subsection{Metrics families} \label{sec:metrics_fam}

\subsubsection{Non-spinning
subclass of the Johannsen-Psaltis family}

\begin{figure*}[htbp]
    \centering
    \begin{tabular}{ccc}
        \hspace{-4mm}
        {\includegraphics[scale=0.43,trim=1.5cm 0 1.5cm 0]{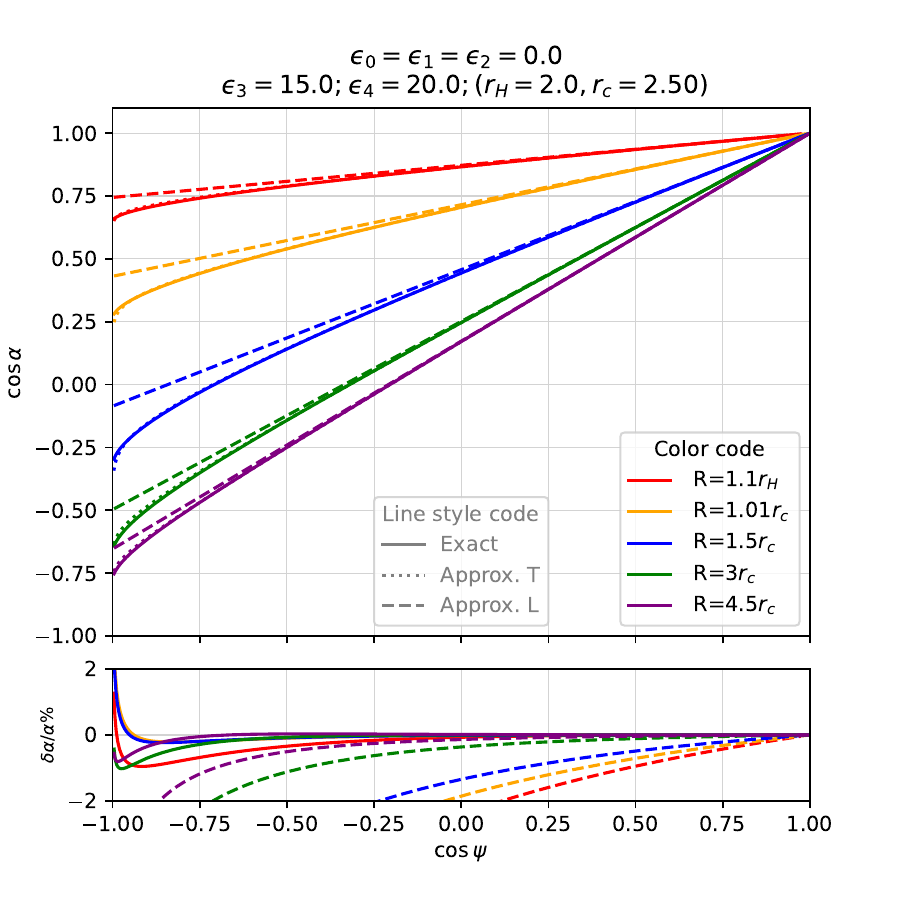}} & 
        \hspace{0.8cm}
        {\includegraphics[scale=0.43,trim=1.5cm 0 1.5cm 0]{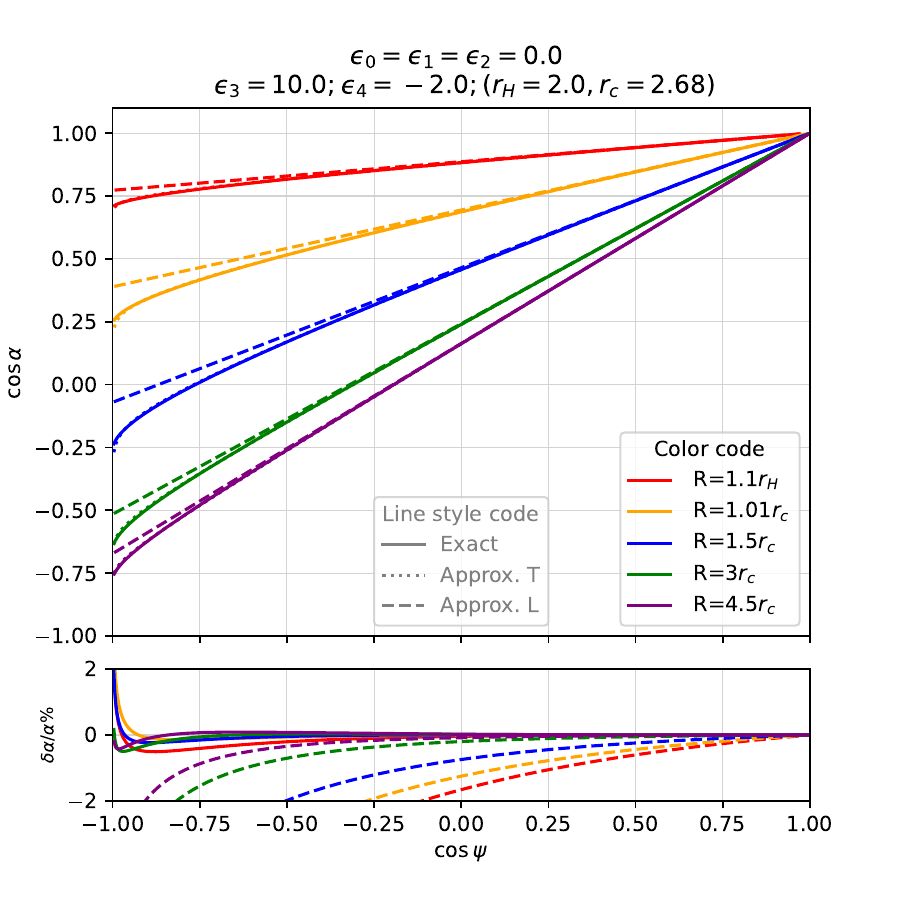}} &
        \hspace{0.8cm}
        {\includegraphics[scale=0.43,trim=1.5cm 0 1.5cm 0]{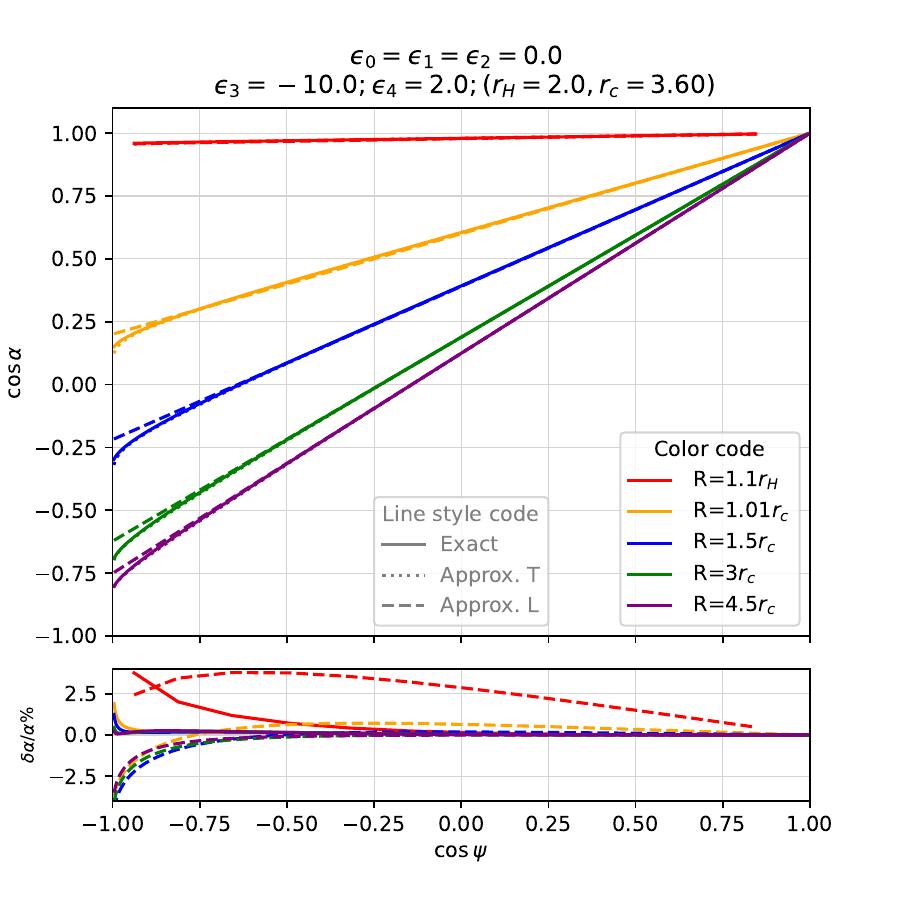}}
    \end{tabular}
    \caption{This figure compares the numerical integration of the exact relation between $\cos\psi$ and $\cos\alpha$ (labeled `Exact') with the analytical approximations given by Eqs.~\eqref{eq:our_aprox_full} (`Approx.~T') and~\eqref{eq:our_aprox_belo} (`Approx.~L'). { As explained in the main text, the numerical integration is performed using the two expressions \eqref{eq:belo-ap1} and \eqref{eq:psi_p}, as appropriate, and joined at the periaxis where $\cos\alpha = 0$ and $\psi=\psi_{\text{max}}$. For the analytical approximation, Eqs.~\eqref{eq:our_aprox_full} or \eqref{eq:our_aprox_belo} are invoked directly over the entire range.} The values $r_H$ and $r_c$ denote the radial coordinates of the event horizon and the photon sphere, respectively, for the Johannsen--Psaltis metric. The bottom panel shows the relative error $\delta\alpha/\alpha\,[\%]$. Solid lines indicate the error for `Approx.~T' (vs `Exact'), while dashed lines indicate the error for `Approx.~L' (vs `Exact').
}
    \label{fig:johan_cospsi}
\end{figure*}

To analyze light curves in the strong-field regime for spacetimes characterized by multiple parameters, the Johannsen–Psaltis (JP) metric~\cite{PhysRevD.83.124015} provides a convenient parametrization. This family describes stationary, axisymmetric, and asymptotically flat Kerr-like black holes that are regular everywhere outside the event horizon. The original motivation for introducing this metric was to furnish a general framework for strong-field tests of the no-hair theorem in the electromagnetic band, exploiting the properties of accretion flows around astrophysical black holes with arbitrary spin~\cite{Johannsen:2012ng,Bambi:2015ldr,Khodadi:2021gbc}. Since its introduction, the JP metric has been widely employed to model departures from Kerr and to confront a variety of geometries and observables. 

A non-exhaustive list of applications of this metric includes studies of the broad K$\alpha$ iron line; modeling of black hole shadows and their comparison with observations; analyses of how deviations from the Kerr geometry affect non-equatorial photon orbits; and studies of gravitational waves from non-Kerr black holes~\cite{Bambi:2012at,Bambi:2015kza,Glampedakis:2018blj,Younsi:2016azx,Olmo:2025ctf,Carson:2020iik,Dey:2022pmv}.
 In this work, the focus is exclusively on the non-spinning case, so that the metric functions $A(r)$ and $B(r)$ of Eq.~\eqref{eq:metric_spheric} are written as
\begin{eqnarray} \label{eq:johan_metric}
    A(r)&=&f(r)(1+h(r)), \nonumber \\
    B(r)&=&\frac{1+h(r)}{f(r)},
\end{eqnarray}
where
\begin{equation} \label{eq:johan_metric_func}
    f(r)=1-\frac{2M}{r},\quad h(r)=\sum_{k=0}^{\infty} \epsilon_k\left(\frac{M}{r}\right)^k,
\end{equation}
with $\epsilon_k$ dimensionless parameters describing the different metrics and $M$ associated to the mass of the central object \cite{PhysRevD.83.124015}. By substitution of $r=R/u'$ into Eqs.~\eqref{eq:int-main} with $i=1,2,3$, is obtained:
\begin{widetext}
\begin{subequations} \label{eq:johan_aprox_gral}
\begin{align} 
    \mathcal{I}_1=&1+\sum_{k=0}^{\infty}\frac{\epsilon_k}{k+1}\left(\frac{M}{R}\right)^k, \label{eq:johan_aprox_1} \\
    \mathcal{I}_2
    =&\frac{1}{3}-\frac{M}{2R}+2 \sum_{k=0}^{\infty} \epsilon_k\left(\frac{M}{R}\right)^k\left(\frac{1}{k+3}-\frac{2M}{R(k+4)}\right) +\sum_{k=0}^{\infty} \left[\left( \sum_{i=0}^{k} \epsilon_i \epsilon_{k-i} \right)  \left(\frac{M}{R}\right)^k\left(\frac{1}{k+3}-\frac{2M}{R(k+4)}\right)\right], \label{eq:johan_aprox_2}\\
    \mathcal{I}_3
    =&\frac{1}{5}-\frac{2M}{3R}+\frac{4M^2}{7R^2}+3\sum_{k=0}^{\infty} \epsilon_k\left(\frac{M}{R}\right)^k\left(\frac{1}{k+5}-\frac{4M}{R(k+6)} + \frac{4M^2}{R^2(k+7)}\right) \nonumber \\
    +&3\sum_{k=0}^{\infty} \left[\left( \sum_{i=0}^{k} \epsilon_i \epsilon_{k-i} \right)  \left(\frac{M}{R}\right)^k\left(\frac{1}{k+5}-\frac{4M}{R(k+6)} + \frac{4M^2}{R^2(k+7)}\right)\right] \nonumber \\
    +&\sum_{k=0}^{\infty} \left\{\left[ \sum_{i=0}^{k} \left(\sum_{j=0}^{i}\epsilon_j \epsilon_{i-j}\right)\epsilon_{k-i} \right]  \left(\frac{M}{R}\right)^k\left(\frac{1}{k+5}-\frac{4M}{R(k+6)} + \frac{4M^2}{R^2(k+7)}\right)\right\}.    \label{eq:johan_aprox_3}
\end{align}    
\end{subequations}
\end{widetext}

For the subsequent numerical calculations, terms up to the fourth order in $k$ in the function $h(r)$, as defined in Eq.~\eqref{eq:johan_metric_func}, are retained.
Since the metrics are required to be asymptotically flat and to satisfy the constraints of the parametrized post-Newtonian (PPN) formalism~\cite{Will:2005va}, the first three parameters must be set to zero~\cite{PhysRevD.83.124015}, i.e., $\epsilon_i = 0$ for $i = 0, 1, 2$. In that case, $M$ coincides with the Arnowitt–Deser–Misner (ADM) mass of the black hole. In what follows, the approximation is tested for different values of $\epsilon_3$ and $\epsilon_4$, with $M$ fixed to $1$.
Figure~\ref{fig:johan_cospsi} shows the relation between $\cos\psi$ and $\cos\alpha$ obtained from our analytical approximations, in direct comparison with the results of the numerical integration of \eqref{eq:belo-ap1} and~\eqref{eq:psi_p}, which must be inverted to express $\alpha$ as a function of $\psi$. The curves corresponding to Eqs.~\eqref{eq:our_aprox_full} and~\eqref{eq:our_aprox_belo} represent the two analytical approximations under evaluation. The lower panel displays the relative percentage error in $\alpha$, illustrating the deviation of each analytical model from the numerical results.
As shown in Fig.~\ref{fig:johan_cospsi}, Eq.~\eqref{eq:our_aprox_full} is virtually indistinguishable from the numerical result for all values of $R$ and parameter choices considered, maintaining a relative error below $3\%$ even in the vicinity of the event horizon. 
In contrast, Eq.~\eqref{eq:our_aprox_belo} gradually loses accuracy as $\psi \rightarrow \pi$; however, for $R > 1.5\,r_c$ it still provides a good approximation, with the relative error remaining below $2\%$ for $\cos\psi > 0.25$.

\begin{figure*}[htbp]
    \centering
    \begin{tabular}{ccc}
        \hspace{-4mm}
        {\includegraphics[scale=0.43,trim=1.5cm 0 1.5cm 0]{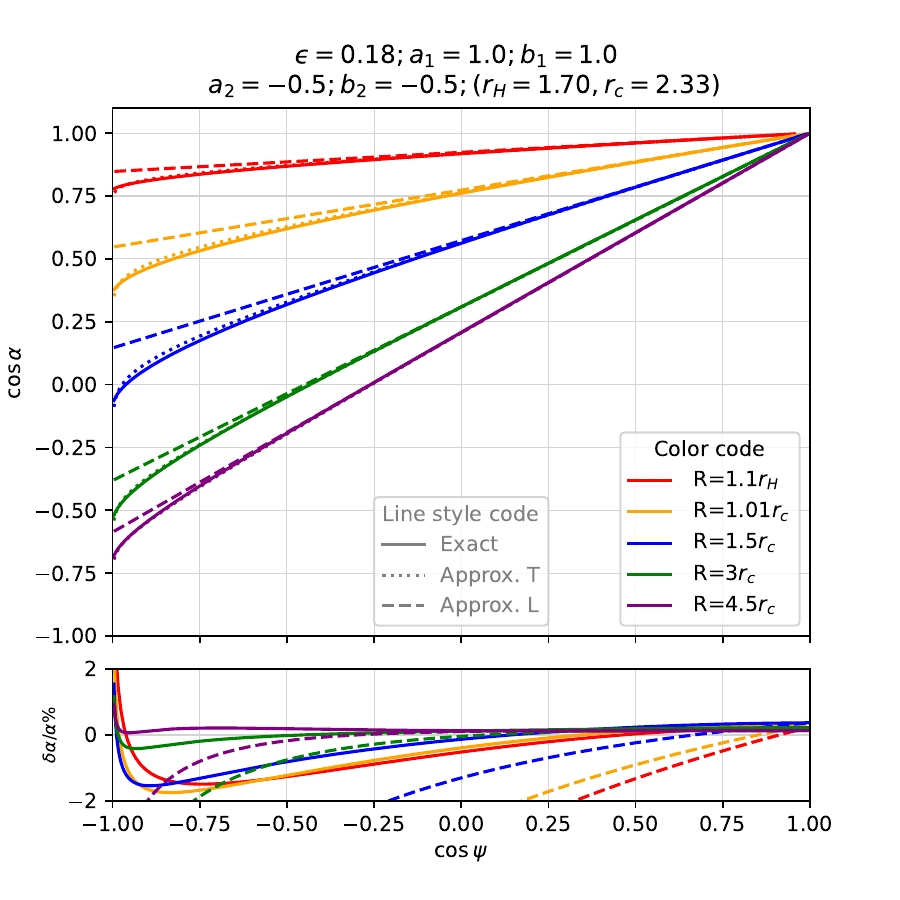}} & 
        \hspace{0.8cm}
        {\includegraphics[scale=0.43,trim=1.5cm 0 1.5cm 0]{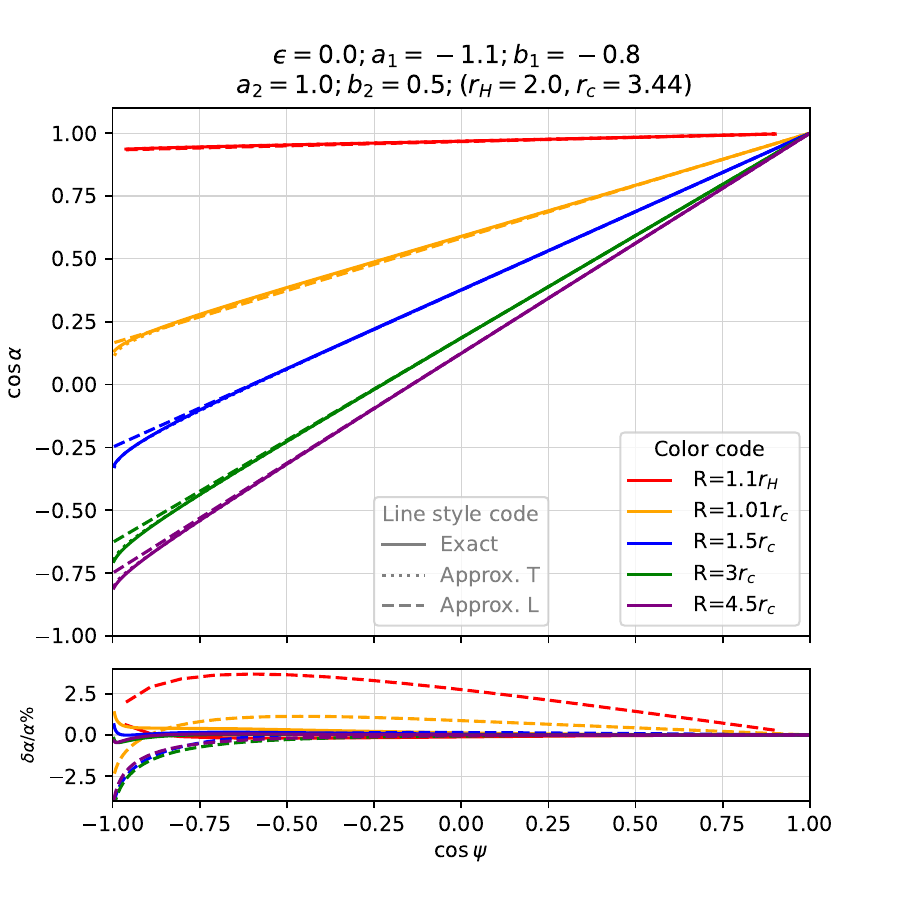}} &
        \hspace{0.8cm}
        {\includegraphics[scale=0.43,trim=1.5cm 0 1.5cm 0]{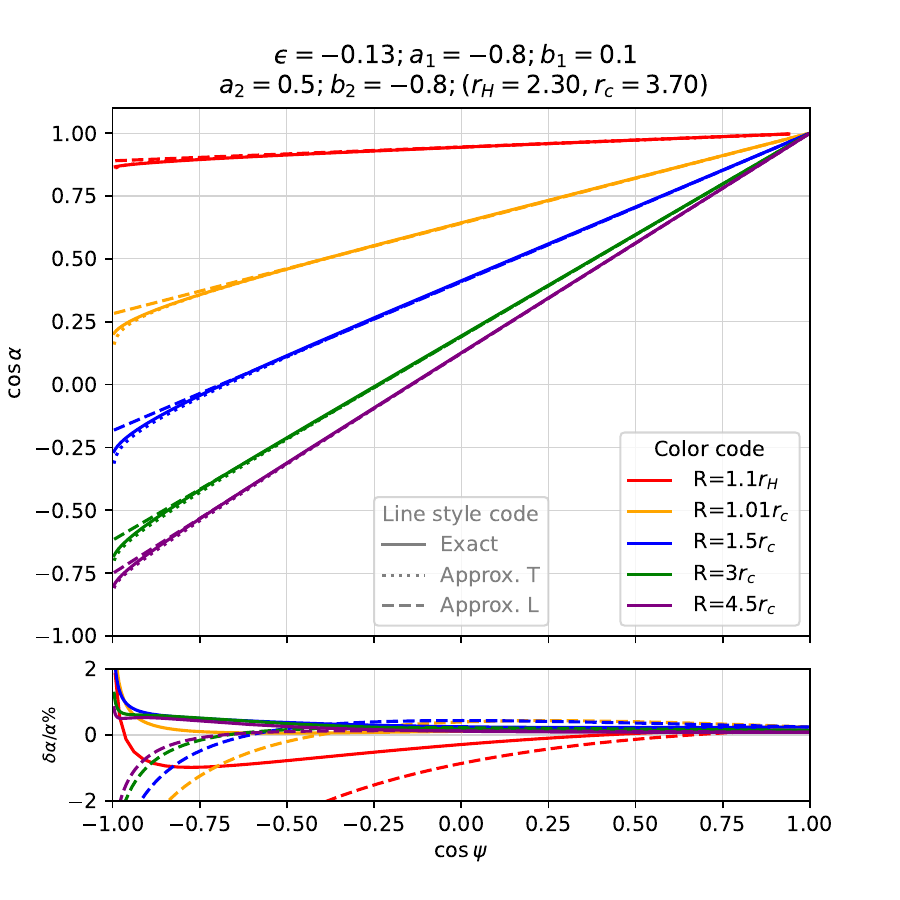}}
    \end{tabular}
    \caption{Similar to Figure \ref{fig:johan_cospsi}, this plot compares the numerical integration of the exact relation $\cos \psi \text{ vs}\cos \alpha$ with the approximations \eqref{eq:our_aprox_full} \eqref{eq:our_aprox_belo} for the Rezzolla-Zhidenko metric.}
    \label{fig:rezzo_cospsi}
\end{figure*}

\subsubsection{Rezzolla-Zhidenko family}

Another metric with a high parameter dependence and widely used in the literature is the Rezzolla-Zhidenko \cite{Rezzolla:2014mua} (RZ) metric, commonly used to test the strong field regime through black hole shadows and strong gravitational lensing\cite {Younsi_2016,Konoplya_2019,Volkel:2020xlc,Mizuno:2018lxz,Feleppa:2025ejh}, as well as parameterization of metrics for alternative theories of gravity \cite{DeLaurentis:2017dny,Kokkotas:2017ymc,Hennigar:2018hza} and, more recently, the study of quasi-normal modes and modifications of the black hole radiation spectrum \cite{Dubinsky:2024nzo,Bolokhov:2025uxz,Siqueira:2025lww}. In contrast to similar approaches that rely on a Taylor expansion of the metric functions $A(r)$ and $B(r)$ in powers of $M/r$ (with $M$ the ADM mass), the authors employ a continued-fraction expansion of these functions. This representation offers improved convergence behavior and enables an accurate approximation of a nontrivial number of known metric solutions from various alternative theories of gravity with only a small set of coefficients. As shown by the authors, determining these coefficients from observations of near-horizon phenomena provides an effective way to constrain and compare different metric theories of gravity.
{ These functions depend on an infinite set of dimensionless parameters: $\epsilon = (2M - r_H)/r_H$, which measures deviations from the Schwarzschild radius; $a_0, a_1, a_2, \ldots$, which represent corrections to the Schwarzschild expression for $A(r)$; and $b_0, b_1, b_2, \ldots$, which represent corrections to the Schwarzschild expression for $B(r)$.} As in the JP metric, both $a_0$ and $b_0$ are negligible within the PPN framework. Remarkably, Ref.~\cite{PhysRevD.101.124004} showed that only a small set of parameters is needed to probe the strong-field regime. In particular, truncating the continued-fraction expansion of $A(r)$ and $B(r)$ at second order leaves $\epsilon$, $a_1$, $a_2$, $b_1$, and $b_2$ as free parameters, which suffice to describe astrophysical observables associated with near-horizon processes. As before, the ADM mass is fixed to unity.
Under these considerations, the metric functions can be written as
\begin{eqnarray} \label{eq:rezzo_metric}
\begin{split}
    A(r)=&\left(1-\frac{r_H}{r}\right)\left(1-\frac{r_H\epsilon}{r}-\frac{r_H^2\epsilon}{r^2}\right.\\&\left.+\frac{r_H^3}{r^2}\frac{a_1}{r+a_2(r-r_H)}\right), \nonumber \\
    B(r)=&\frac{1}{A(r)}\left(1+\frac{r_H^2}{r}\frac{b_1}{r+b_2(r-r_H)}\right)^2.
    \end{split}
\end{eqnarray}
Returning to our analytical approximations, for this particular metric we have retained, in the comparison with the numerical integration, only terms up to second order in $y$ in Eq.~\eqref{eq:our_aprox_full}, since the evaluation of $\mathcal{I}_3$ becomes increasingly cumbersome. Therefore, our analytical treatment and subsequent applications include only the first two integrals, $\mathcal{I}_1$ and $\mathcal{I}_2$, as defined in Eqs.~\eqref{eq:int-main}. Nevertheless, as we will show, these approximations preserve high fidelity with respect to the exact integral. According to the above, one obtains
\begin{widetext}
\begin{subequations} \label{eq:rezzo_aprox_gral}
\begin{eqnarray}\label{eq:rezzo_aprox}
    \mathcal{I}_1&=&1 - \frac{b_1 r_H}{2 b_2 R} - \frac{b_1 (1 + b_2) }{b_2^2} - \frac{b_1 (1 + b_2)^2 R }{b_2^3 r_H}\ln\left(1-\frac{b_2 r_H}{(1 + b_2) R}\right), \label{eq:rezzo_aprox_1} \\
    \mathcal{I}_2&=& - \sum_{k=1}^{7}\frac{P_k}{a_2^{8-k}b_2^{8-k}k} \left(\frac{R}{r_H}\right)^{3-k}+ \frac{R^3}{r_H^3(a_2-b_2)}\left\{\frac{a_1}{a_2^8}(1 + a_2)^5 \left[b_1 + a_2^2 (1 + b_1) + a_2(2 b_1 - b_2)\right]\ln\left(1-\frac{a_2r_H}{(1 + a_2) R}\right)\right. \nonumber \\
    &&-\left.\frac{b_1}{b_2^8}(1 + b_2)^4 \left[a_1 (1 + b_2)^3 + (a_2 - b_2) \left((1 + b_2) (1 + 2 b_2) \epsilon - b_2^2\right)\right]\ln\left(1-\frac{b_2r_H}{(1 + b_2) R}\right)\right\}, \label{eq:rezzo_aprox_2} \\
    \text{with} \nonumber \\
    P_k&:=&\frac{a_1  b_1}{b_2 - a_2}\left[b_2^{8-k}(1 + a_2)^{7-k} - a_2^{8-k}(1 + b_2)^{7-k}\right] -(1 - \delta_{k,7}) a_1 a_2 b_2^{8-k} (1 + a_2)^{\max(5-k,0)} \nonumber \\
    &&+ (1 - \delta_{k,5} - \delta_{k,6} - \delta_{k,7})a_2^{8-k} b_1 (1 + b_2)^{\max(4-k,0)} (\epsilon (1 + b_2 )(1 + 2 b_2 ) - b_2^2) \nonumber \\
    &&- \delta_{k,3}a_2^5 b_2^5  + \delta_{k,4}a_2^4 b_2^4 (1 + \epsilon) +\delta_{k,5}a_2^3 b_1 ((1 + 2 b_2)\epsilon - b_2^2) +  \delta_{k,6}a_2^2\epsilon((1 + b_2)b_1 - b_2^2) +  \delta_{k,7}a_2 b_1 \epsilon. \nonumber
\end{eqnarray}
\end{subequations}
\end{widetext}

Figure~\ref{fig:rezzo_cospsi} displays the relation between $\cos\psi$ and $\cos\alpha$ for five emission radii, comparing the numerical integration (solid) with the analytical approximations in Eqs.~\eqref{eq:our_aprox_full} (`Approx.~T', dotted) and~\eqref{eq:our_aprox_belo} (`Approx.~L', dashed). The three panels correspond to distinct choices of the continued--fraction parameters $(\epsilon;a_1,a_2;b_1,b_2)$, with the associated $(r_H,r_c)$ indicated on top. In all cases, `Approx.~T' tracks the numerical curves extremely well across the full range of $\cos\psi$, with a relative error that remains at the level of $\lesssim 2\%$ for all radii considered, including the near--horizon case $R=1.1\,r_H$ and the photon--sphere--proximate case $R=1.01\,r_c$. As expected, both approximations converge to the numerical result in the weak--deflection regime ($\cos\psi\!\to\!1$), where the curves nearly overlap.

The linearized scheme (`Approx.~L') behaves well in the moderate--bending regime, providing an accurate description for $\cos\psi>0$ and $R\gtrsim1.5\,r_c$. Its accuracy degrades systematically as $\psi\to\pi$ (i.e., $\cos\psi\to-1$), particularly for the smallest radii ($R=1.1\,r_H$ and $R=1.01\,r_c$), where the dashed curves depart most visibly from the solid ones. The bottom panels quantify these trends: the error of `Approx.~T' stays within a few percent and shows smooth, parameter--dependent sign changes, whereas `Approx.~L' exhibits larger biases at large deflection but remains competitive for $\cos\psi\gtrsim0$ and larger $R$. Importantly, the qualitative conclusions are robust across the different parameter sets: varying $(\epsilon;a_1,a_2;b_1,b_2)$ does not alter the observed hierarchy in performance between `Approx.~T' and `Approx.~L'.

\subsection{A particular metric} \label{sec:metrics_part}

\subsubsection{Einstein-Maxwell-dilaton-axion metric} \label{sec:metric_EMDA} 

Let us now consider the particular case of the Einstein--Maxwell--dilaton--axion (EMDA) metric, which describes a black hole solution arising in the low-energy limit of string theory~\cite{GIBBONS1982337,GIBBONS1988741,PhysRevD.43.3140}. In this work, we adopt the parametrization introduced in Ref.~\cite{PhysRevLett.74.1276}, setting both the spin and the axion field to zero. Under these assumptions, the line element takes the form
\begin{equation} \label{eq:metric_garcia}
    d\tilde{s}^2=-\tilde{A}(\rho)\,dt^2+\tilde{B}(\rho)\,d\rho^2 + \bigl(\rho^2+2b_{0}\rho\bigr)\,d\Omega^2,
\end{equation}
where
\begin{eqnarray} \label{eq:metric_garcia_prev}
    \tilde{A}(\rho)&=&\frac{\rho-2\mu}{\rho+2b_0}, \nonumber \\
    \tilde{B}(\rho)&=&\frac{\rho+2b_0}{\rho-2\mu}.
\end{eqnarray}
Here, $d\Omega^2$ denotes the metric on the unit 2-sphere, $b_0$ is the dilaton parameter, and $\mu$ is related to the black-hole mass. The radial coordinate $r$ and the ADM mass $M$ are related to the original parameters by
\begin{equation*}
r^2=\rho^2+2b_0\rho, \qquad M=\mu+b_0,
\end{equation*}
so that, after this coordinate transformation, the metric functions can be rewritten as
\begin{eqnarray} \label{eq:emda_metric}
    A(r)&=&\frac{-b_0+\sqrt{b_0^2+r^2}-2\mu}{\,b_0+\sqrt{b_0^2+r^2}\,}, \nonumber \\
    B(r)&=&\frac{1}{A(r)}\,\frac{r^2}{b_0^2+r^2}.
\end{eqnarray}

By substituting the metric functions from Eq.~\eqref{eq:emda_metric} into the integrals defined in Eq.~\eqref{eq:int-main} for $n=1,2,3$, and introducing the dimensionless parameters $a := b_0/R$ and $c := \mu/R$, one obtains
\begin{widetext}
\begin{eqnarray*}\label{eq:emda_aprox}
    \mathcal{I}_1&=& \frac{1}{a}\ln\left(a+\sqrt{1+a^2}\right), \\
    \mathcal{I}_2&=& -\frac{a+c}{2}\left(1-\frac{\sqrt{1+a^2}}{a}\right) + \frac{a+3c}{4a^3}\left[\frac{1}{a}\ln\left(a+\sqrt{1+a^2}\right)- \sqrt{1+a^2}\right], \\
    \mathcal{I}_3&=&-\frac{a+c}{3}\left(2+3a(a+c)\right) + \frac{(a+c)(a+3c)+6c(a+2c)}{16 a^7}\ln\left(a+\sqrt{1+a^2}\right)  \\
    &&+ \frac{\sqrt{1 + a^2}}{48a^6}\left[48a^6(a+c)^2-\frac{1}{3}(15-10a^2+8a^4)(a+3c)^2+\frac{2}{3}a^2(3-2a^2+16a^4)\right]. 
\end{eqnarray*}
\end{widetext}

\begin{figure*}[htbp]
    \centering
    \begin{tabular}{ccc}
        \hspace{-4mm}
        {\includegraphics[scale=0.43,trim=1.5cm 0 1.5cm 0]{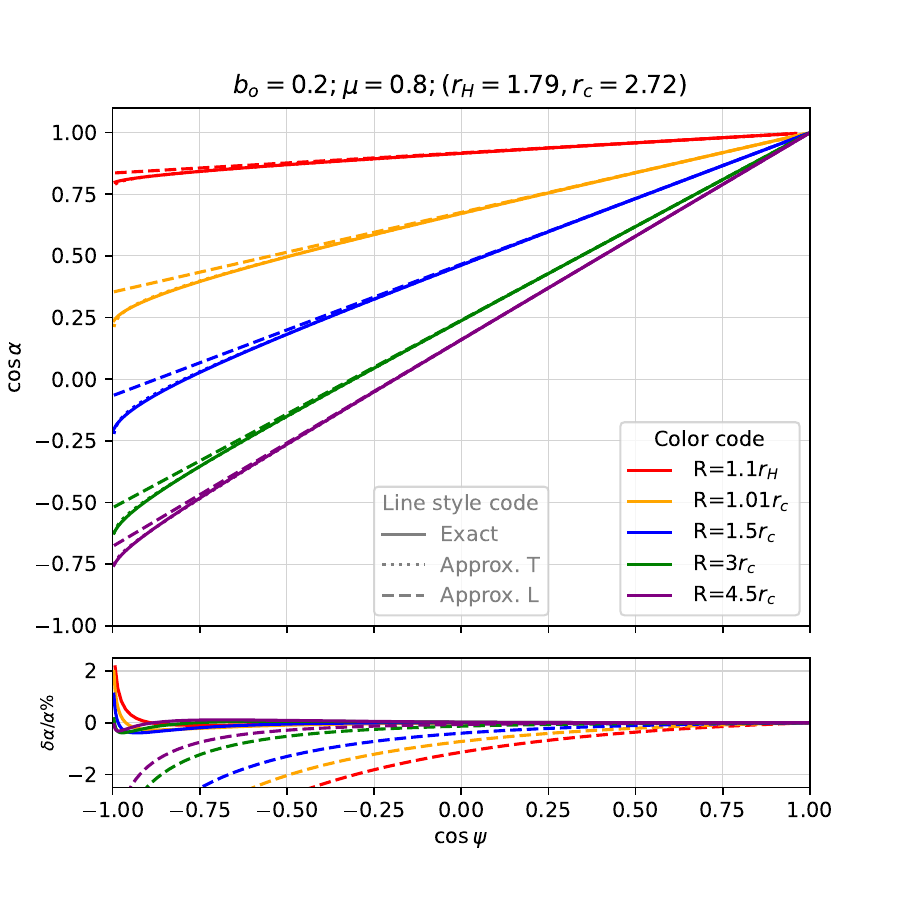}} & 
        \hspace{0.8cm}
        {\includegraphics[scale=0.43,trim=1.5cm 0 1.5cm 0]{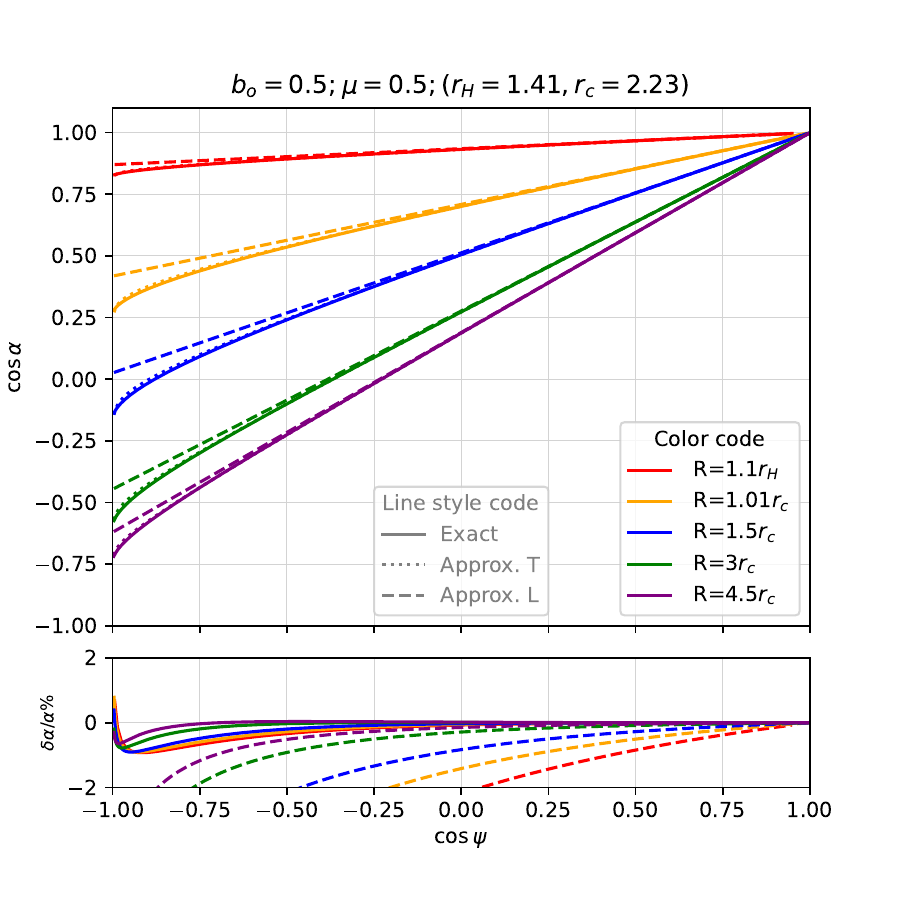}} &
        \hspace{0.8cm}
        {\includegraphics[scale=0.43,trim=1.5cm 0 1.5cm 0]{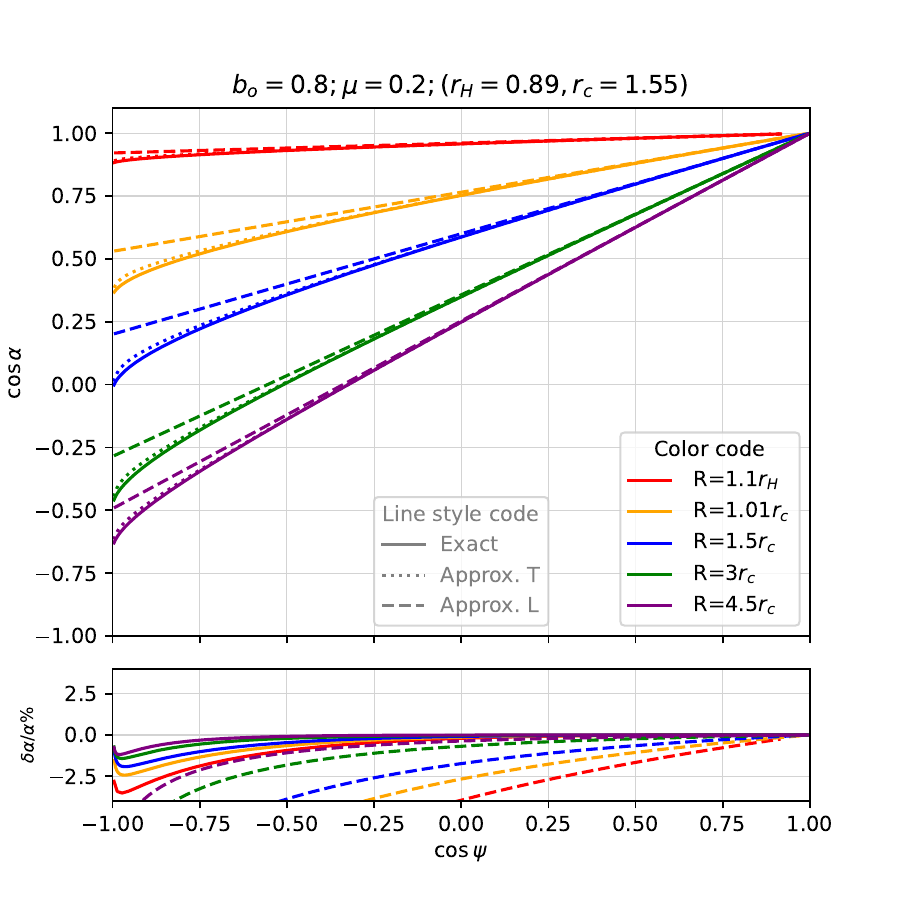}}
    \end{tabular}
    \caption{Similar to Figures \ref{fig:johan_cospsi} and \ref{fig:rezzo_cospsi}, this plot provides the comparison for the Einstein-Maxwell-dilaton-axion metric.}
    \label{fig:emda_cospsi}
\end{figure*}

Figure~\ref{fig:emda_cospsi} illustrates the performance of the analytical approximations given by Eqs.~\eqref{eq:our_aprox_full} and~\eqref{eq:our_aprox_belo} for the Einstein--Maxwell--dilaton--axion metric, considering three representative configurations corresponding to different combinations of the parameters $(b_0,\mu)$ and the associated $(r_H,r_c)$ values. The plots display the relation between $\cos\psi$ and $\cos\alpha$ for several emission radii $R$, together with the relative percentage deviation of $\alpha$ in the lower panels. Across all parameter choices, the full analytical approximation given by Eq.~\eqref{eq:our_aprox_full} (Approx.~T) shows excellent agreement with the exact numerical integration. The discrepancy remains within $|\delta\alpha|/\alpha < 3\%$ over the entire domain of $\psi$, even for the smallest radii considered; namely, $R = 1.1r_H$ and $R = 1.01r_c$; where strong deflection effects dominate. This shows that the approximation maintains high accuracy even in the deep strong-field regime near the event horizon. The linearized expression derived from Eq.~\eqref{eq:our_aprox_belo} (`Approx.~L') also performs well for moderate deflections ($\cos\psi \gtrsim 0$), particularly at larger radii ($R \ge 1.5r_c$), but gradually loses precision as $\psi \to \pi$ and $\cos\psi \to -1$. This degradation is more evident in the rightmost panel, where the dilaton parameter dominates ($b_0 > \mu$), leading to a more compact horizon and stronger curvature gradients. In contrast, in the leftmost configuration ($b_0 < \mu$), both approximations remain nearly indistinguishable from the numerical result except for very large bending angles. As in the previous cases, the lower panels quantify these trends. For all three configurations, `Approx.~T' maintains a smooth and nearly symmetric error distribution around $\cos\psi = 0$, indicating robustness under parameter variations. `Approx.~L', while showing larger residuals near $\cos\psi \approx -1$, still provides a reasonable description in the weak-to-intermediate deflection regime, making it suitable for analytical estimates where computational simplicity is preferred.
Overall, these results show that the generalized analytical formulas accurately reproduces light propagation in the EMDA spacetime, even for extreme parameter values where deviations from Schwarzschild geometry are significant.

\section{Some simple applications} \label{sec:applications}
\subsection{Images of isoradial curves of accretion disks} \label{sec:isoradial_curves}
As a first application (and an additional validation) we examine how the analytical formulas derived above can be used to reconstruct images of equatorial isoradial curves, i.e., {curves of fixed $r$} lying in the plane $\theta=\pi/2$. These isoradial curves can be regarded as belonging to a thin accretion disk lying on the equatorial plane. In this context, the $\alpha$--$\psi$ relations provide a fully analytical mapping between the emission radius and the observed direction for direct rays, enabling efficient image reconstruction without the need for full numerical ray tracing.

\begin{figure} [h!]
    \centering
    \includegraphics[scale=0.43]{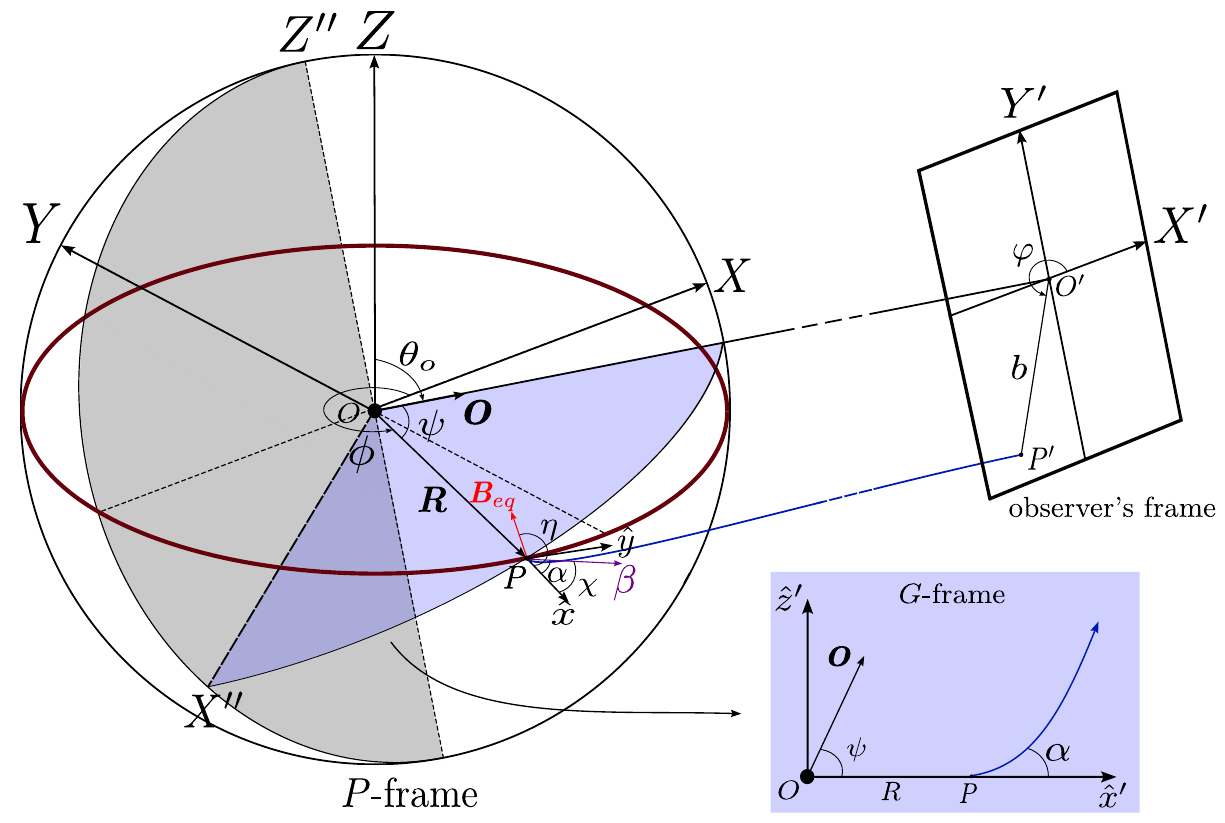}
    \caption{{Figure reproduced from Ref.~\cite{Claros:2024atw}, Fig.~7.} The geometry of the different coordinate frames is illustrated. The orbit of fluid element $P$ at radius $r = R$ (board), the plane containing the geodesic trajectory (blue), and the plane of the distant observer with its corresponding projection (gray) are represented. Vectors used for the study of polarization in Sec.~\ref{sec:qu_diagrams} are also shown: the fluid velocity (purple) and the equatorial projection of the magnetic field (red).}
    \label{fig:gralframe}
\end{figure}

Following the methodology described in Refs.~\cite{Claros:2024atw,Lumi:1979}, let us consider a point $P$ on an accretion disk located on the equatorial $XY$-plane of a sphere with radius $R$, centered at the origin $O$ of the coordinate system $(X,Y,Z)$, where a black hole (or another compact object) is assumed to be positioned. At this point $P$, a fluid element emits radiation and has local coordinates $\vb*{R}=R(\cos\phi,\sin\phi,0)$. The angle $\theta_o$ denotes the inclination of the disk with respect to a distant observer equipped with a camera centered at $O'$, whose local coordinate system is $(X',Y')$. The unit vector $\vb*{o}=(0,-\sin\theta_o,\cos\theta_o)$ represents the direction connecting $O$ and $O'$. The geometry shown in Fig.~\ref{fig:gralframe} is completed by the following relations: $\overline{OX}$ is parallel to $\overline{O'X'}$, and $\overline{OZ''}$ is parallel to $\overline{O'Y'}$, implying that the gray plane is parallel to the observer’s image plane. Furthermore, $\overline{OX''}$ is parallel to $\overline{O'P'}$, and $\overline{OO'}$ is perpendicular to the observer’s frame. As a consequence, the photon trajectory lies entirely within the blue plane, as illustrated in the bottom-right panel of Fig.~\ref{fig:gralframe}. The remaining geometric elements shown in Fig.~\ref{fig:gralframe} are discussed in Sec.~\ref{sec:qu_diagrams}.

To analyze the isoradial images, consider the emission point $P$ and its image $P'$ on the observer’s screen, with Cartesian coordinates $(X'_{P'},Y'_{P'})$ or, equivalently, polar coordinates $(b,\varphi)$ defined by
\begin{equation}\label{eq:polar_coord}
    X'_{P'} = b\cos\varphi, \qquad Y'_{P'} = b\sin\varphi .
\end{equation}
Recalling that $x = 1 - \cos\alpha$, the impact parameter $b$, defined in Eq.~\eqref{eq:impact_parameter}, can be written as a function of the emission radius and $x$ as:
\begin{equation}\label{eq:bgenx}
    b(R,x)=\frac{R}{\sqrt{A(R)}}\sqrt{1-(1-x)^2},
\end{equation}
where $x$ is expressed as a function of $\psi$ through the analytical approximations given in Eqs.~\eqref{eq:our_aprox_full} and~\eqref{eq:our_aprox_belo} derived above or through the numerical integration of \eqref{eq:psi_p} and \eqref{eq:belo-ap1}.
 Moreover, from the inner product between the unit vectors $\vb*{o}$ and $\vb*{R}/|\vb*{R}|$, we obtain a relation between $\psi$ and $\phi$ for a given orientation $\theta_o$ of the camera,
\begin{equation} \label{eq:cospsi_eq}
    \cos\psi=-\sin\theta_{o}\sin\phi.   
\end{equation}

With these relations, $b$ can be written as $b=b(R,\phi)$. Since our goal is to establish a direct mapping between $(R,\phi)$ and $(b,\varphi)$-or, equivalently, between $P$ and $P'$-it is sufficient to determine a relation linking $\varphi$ and $\phi$. This relation is given by (for a concise derivation, see Appendix A of Ref.~\cite{Claros:2024atw}):
\begin{eqnarray}
    \cos\varphi&=&\frac{\cos\phi}{\sqrt{1-\sin^2\theta_o\sin^2\phi}},\label{cosvarp}\\
    \sin\varphi&=&\frac{\sin\phi\cos\theta_{o}}{\sqrt{1-\sin^2\theta_o\sin^2\phi}}.\label{sinvarp}
\end{eqnarray}

Together, all these relations define a direct mapping from the emission point $P$ with local coordinates $(R,\phi)$ to the observation point $P'$ with local coordinates $(b,\varphi)$.

\begin{figure*}[htbp]
    \centering
    \begin{tabular}{ccc}
        \hspace{-6mm}
        {\includegraphics[scale=0.3,trim=2cm 1.65cm 1.24cm 1.7cm, clip]{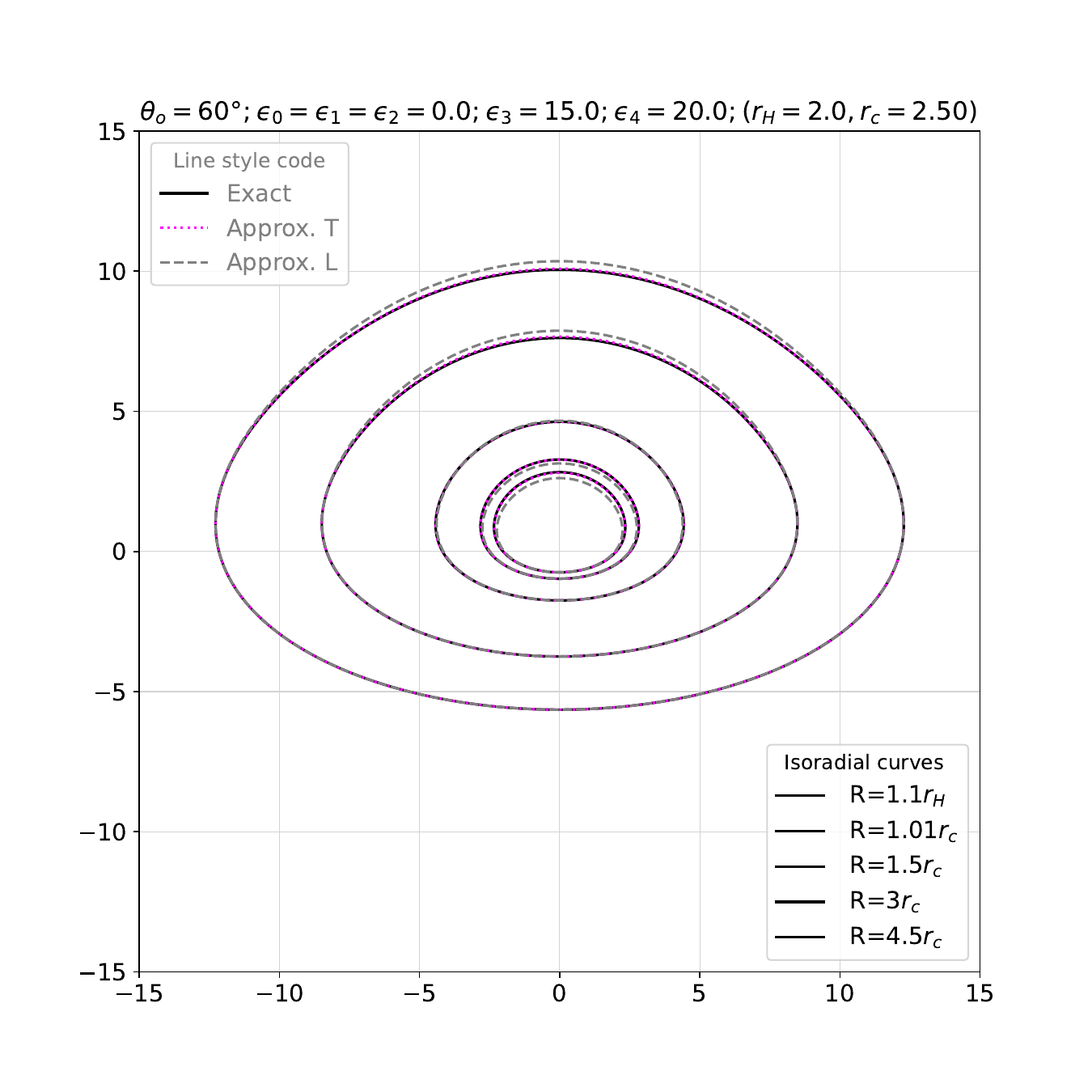}} &
        \hspace{-0.2cm}
        {\includegraphics[scale=0.3,trim=2cm 1.65cm 1.24cm 1.7cm, clip]{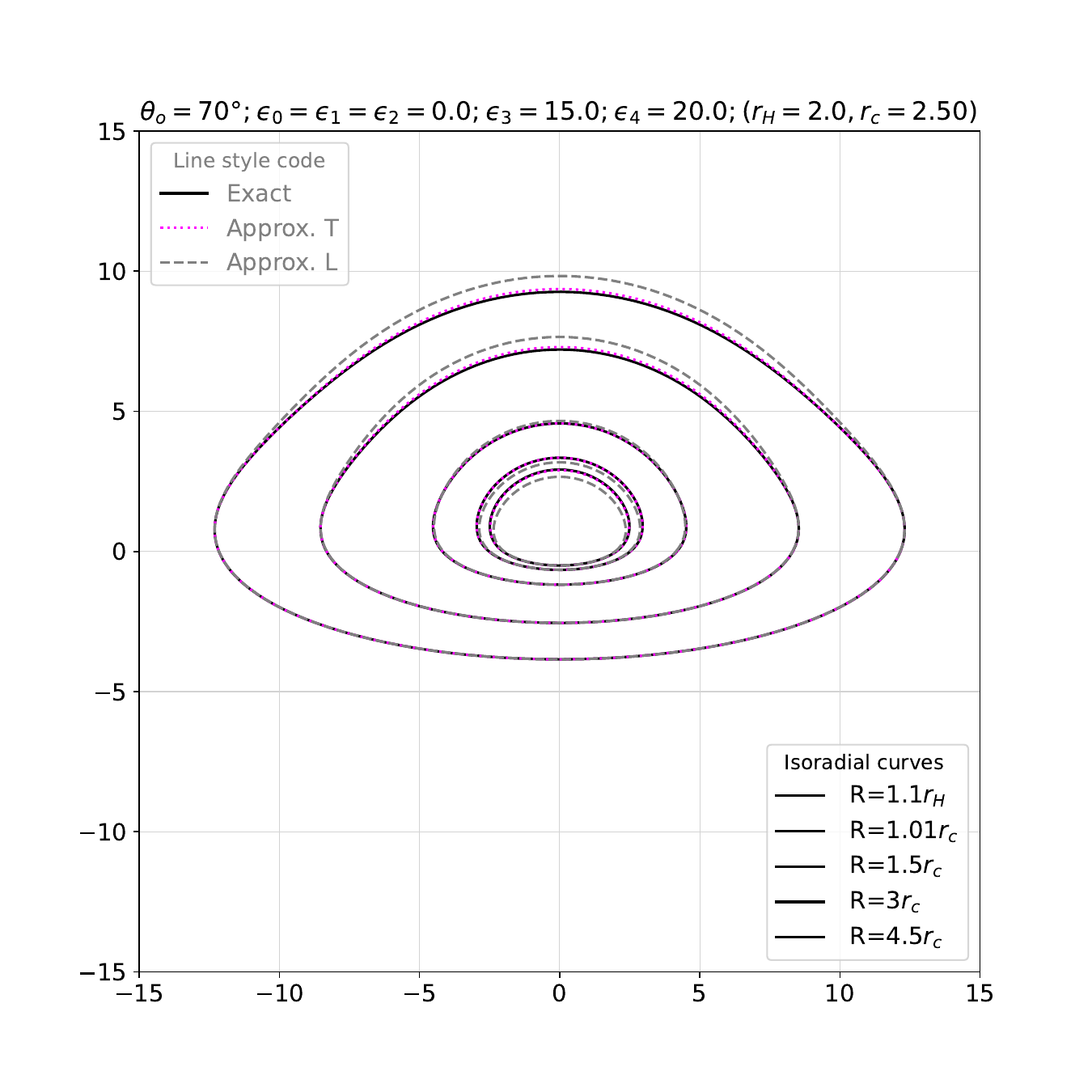}} &
        \hspace{-0.2cm}
        {\includegraphics[scale=0.3,trim=2cm 1.65cm 1.24cm 1.7cm, clip]{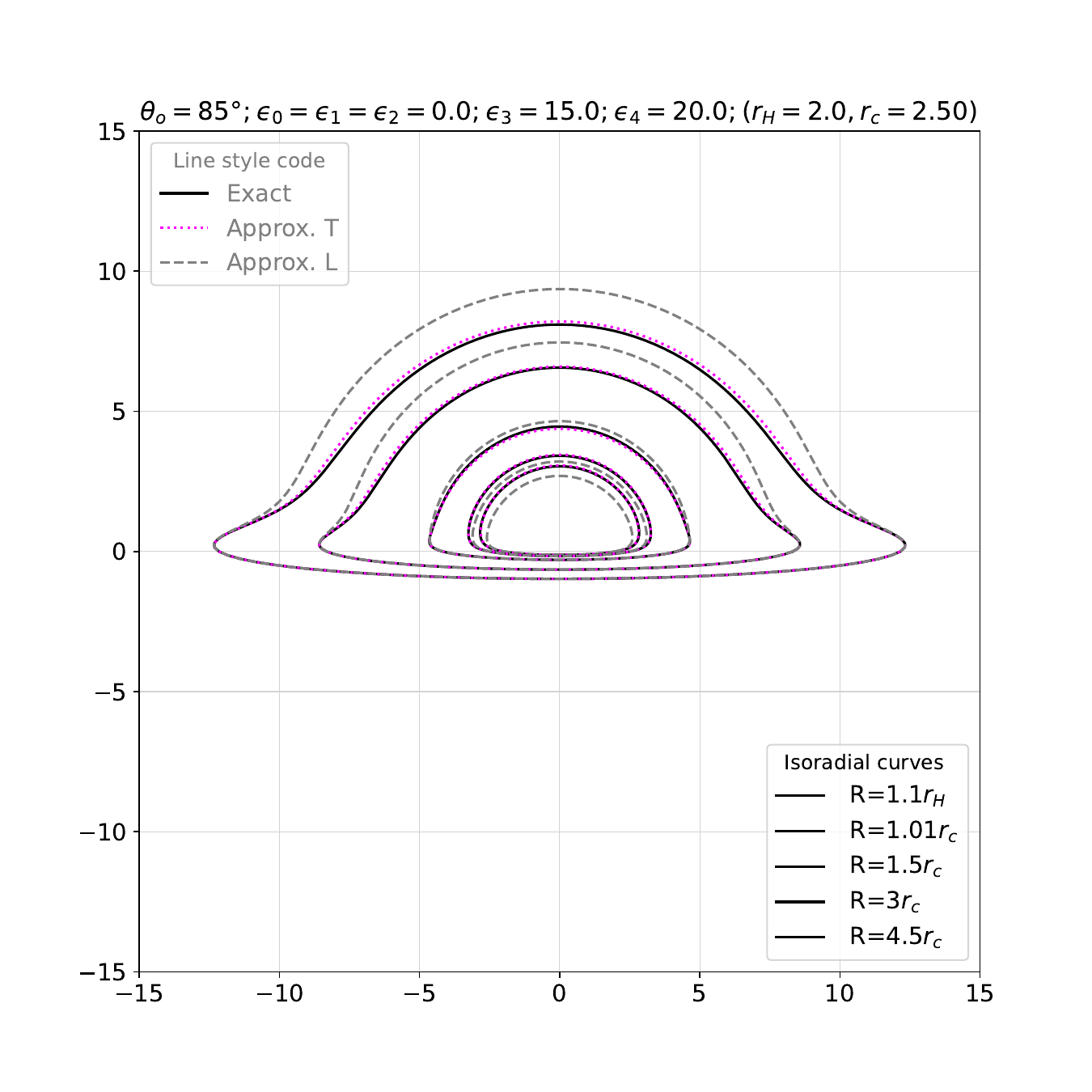}} \\
        \hspace{-6mm}
        {\includegraphics[scale=0.3,trim=2cm 1.65cm 1.24cm 1.7cm, clip]{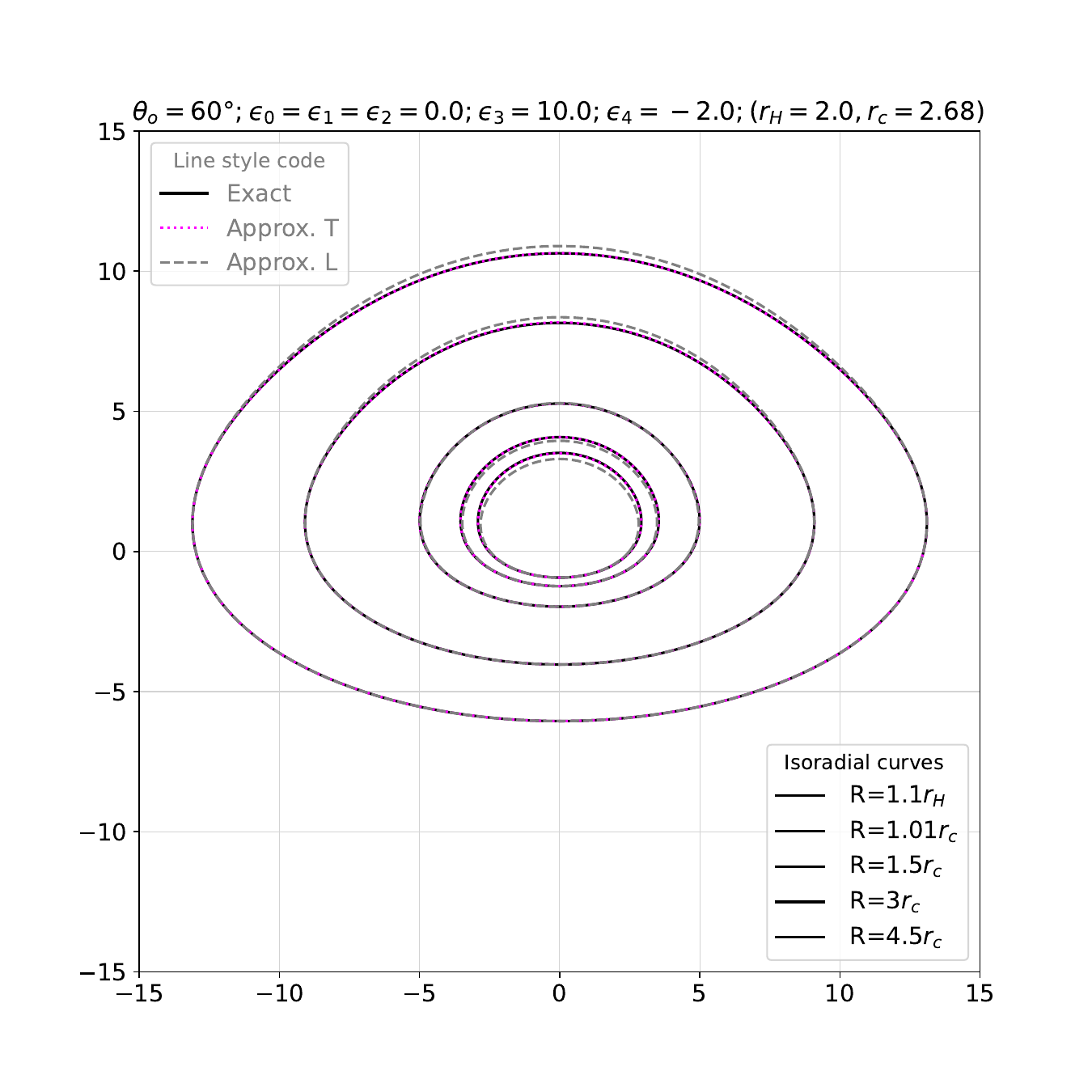}} & 
        \hspace{-0.2cm}
        {\includegraphics[scale=0.3,trim=2cm 1.65cm 1.24cm 1.7cm, clip]{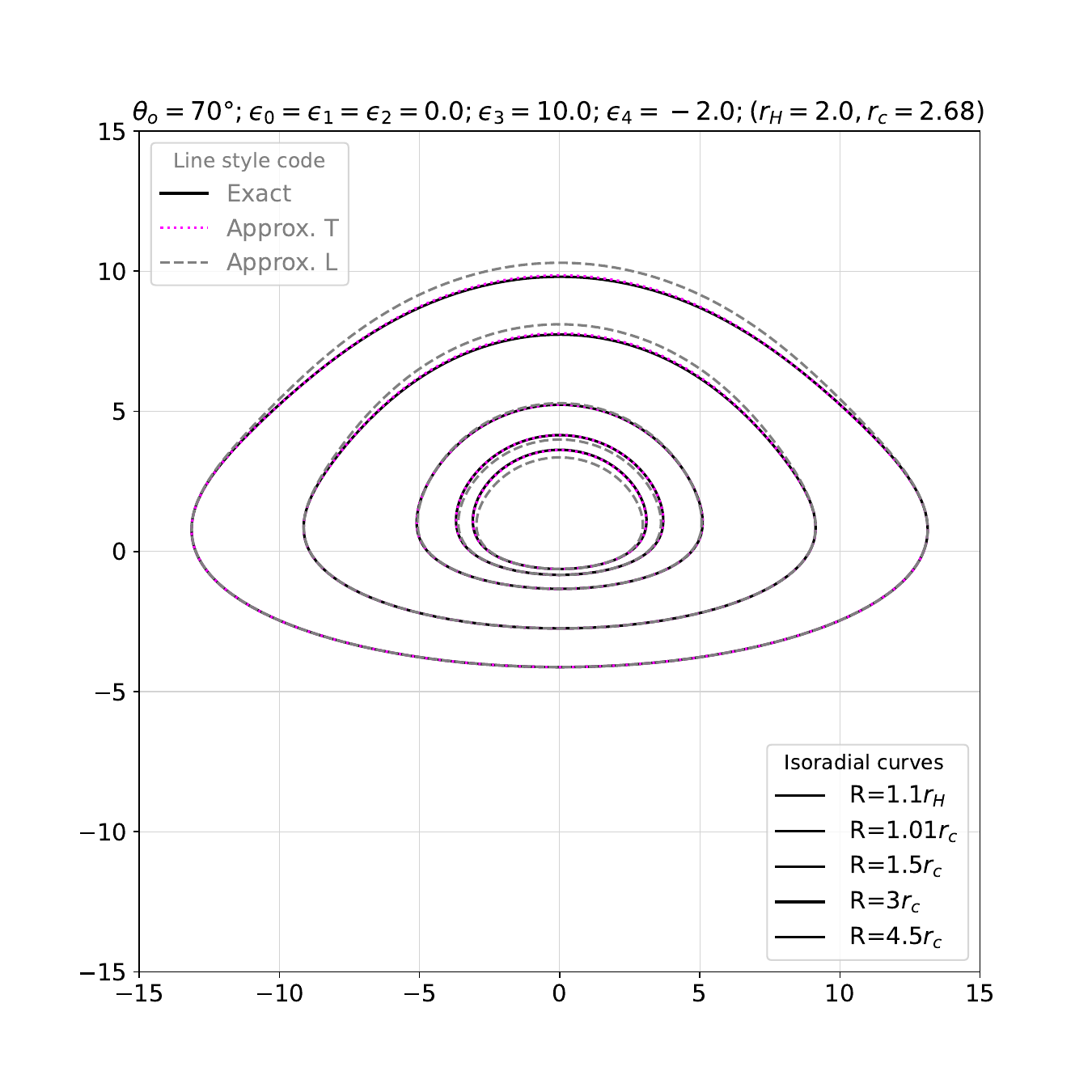}} &  
        \hspace{-0.2cm}
        {\includegraphics[scale=0.3,trim=2cm 1.65cm 1.24cm 1.7cm, clip]{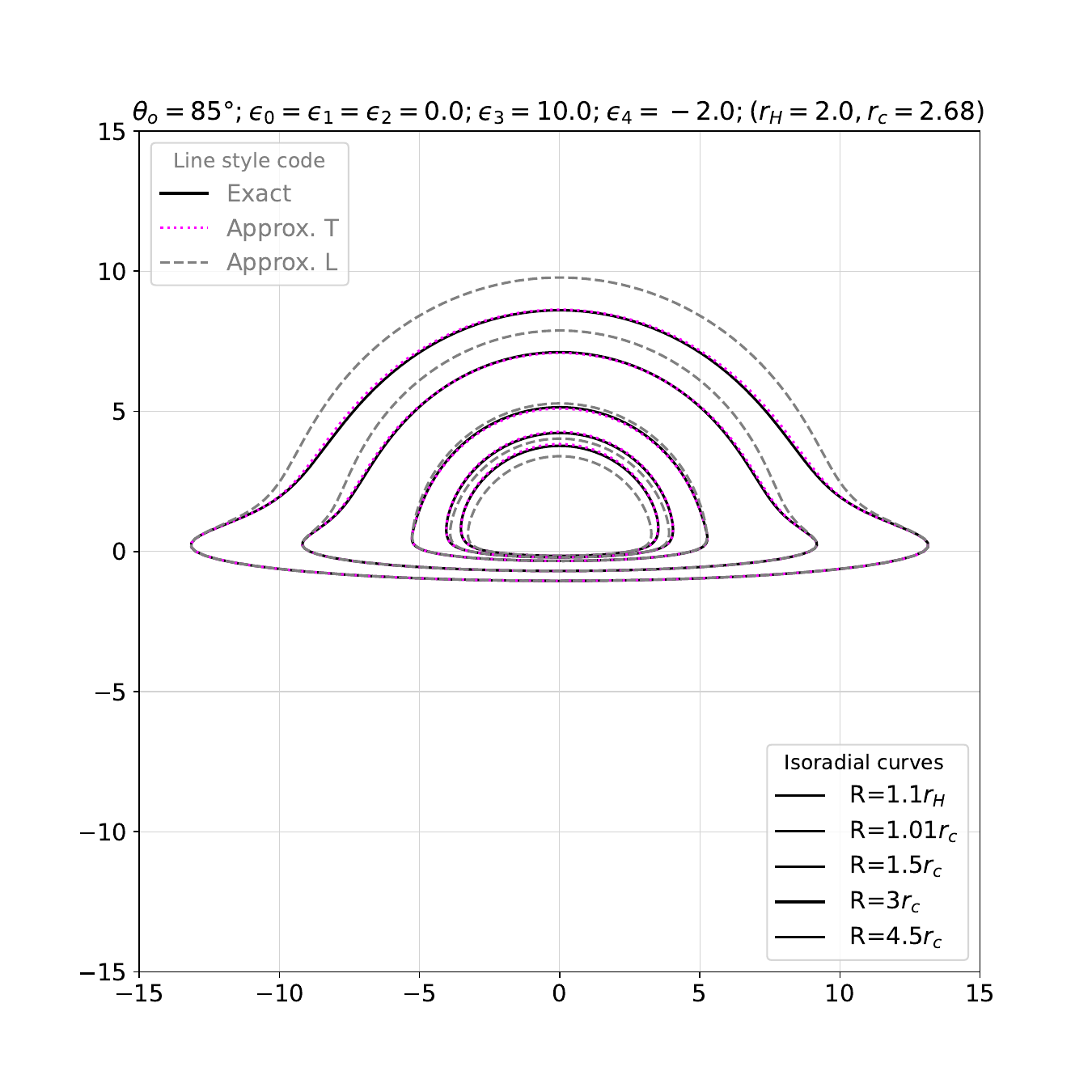}}
    \end{tabular}
    \caption{This figure compares isoradial curves ($R=\text{constant}$) for the Johannsen-Psaltis metric. Two sets of non-zero parameters are shown: $\epsilon_3=15.0, \epsilon_4=20.0$ (top row); $\epsilon_3=10.0, \epsilon_4=-2.0$ (bottom row). Each set is plotted for three observer inclinations: $\theta_o = 60^{\circ}$, $70^{\circ}$ and $85^{\circ}$. The exact curves, calculated numerically (solid line) are compared with two analytical approximations: `Approx. T' (dotted line) and `Approx. L' (dashed line).}
    \label{fig:johan_isorad}
\end{figure*}

To test the accuracy and robustness of our analytical approximations, Fig.~\ref{fig:johan_isorad} displays images of isoradial curves ($R=\text{constant}$) computed for the JP metric under different parameter configurations and viewing inclinations. Two representative sets of non-zero deformation parameters are considered: $(\epsilon_3,\epsilon_4)=(15,20)$ (top row) and $(\epsilon_3,\epsilon_4)=(10,-2)$ (bottom row). Each configuration is evaluated for three observer inclination angles, $\theta_o = 60^{\circ}$, $70^{\circ}$, and $85^{\circ}$. The curves obtained from exact numerical integration (solid lines) are compared with the analytical approximations~\eqref{eq:our_aprox_full} (`Approx.~T', dotted line) and~\eqref{eq:our_aprox_belo} (`Approx.~L', dashed line). The same set of radii used in Sec.~\ref{Sec:Testing} is adopted here.

As the figure shows, the `Approx.~T' curves are in excellent agreement with the numerically integrated ones across all configurations. This high level of consistency persists even at large viewing inclinations; up to $\theta_o=85^{\circ}$; and for outer radii as large as $R=4.5r_c$. The resulting isoradial shapes are virtually indistinguishable, with deviations well below the graphical resolution. This confirms that the full approximation~\eqref{eq:our_aprox_full} reliably captures both the geometric deformation induced by the inclination and the metric corrections introduced by non-zero $\epsilon_i$ parameters. The `Approx.~L' curves, while still providing a good qualitative description, begin to show slight discrepancies compared to the exact and `Approx.~T' results, particularly for the largest radii and high-inclination cases. These deviations manifest mainly as small radial displacements and minor distortions on the far side of the accretion disk, where the light rays must pass close to the black hole before reaching the observer, making light-bending effects more pronounced.
Nevertheless, `Approx.~L' remains sufficiently accurate for moderate inclinations ($\theta_o \lesssim 70^{\circ}$) or when only coarse image reconstruction is required.

\begin{figure*}[htbp]
    \centering
    \begin{tabular}{ccc}
        \hspace{-6mm}
        {\includegraphics[scale=0.31,trim={1.34cm 1.65cm 0.2cm 1.47cm},clip]{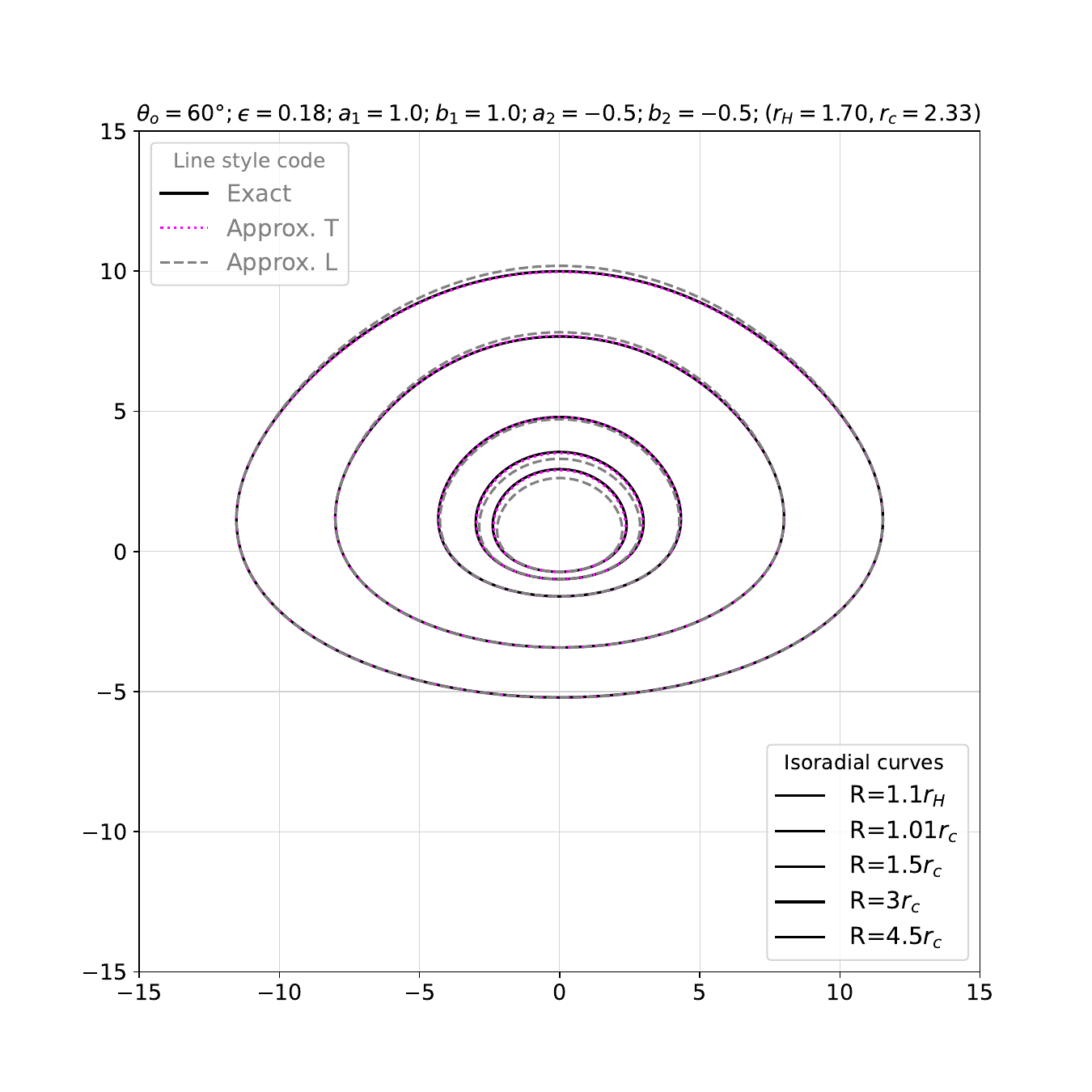}} &
        \hspace{-0.8cm}
        {\includegraphics[scale=0.31,trim={1.7cm 1.65cm 1.26cm 1.47cm},clip]{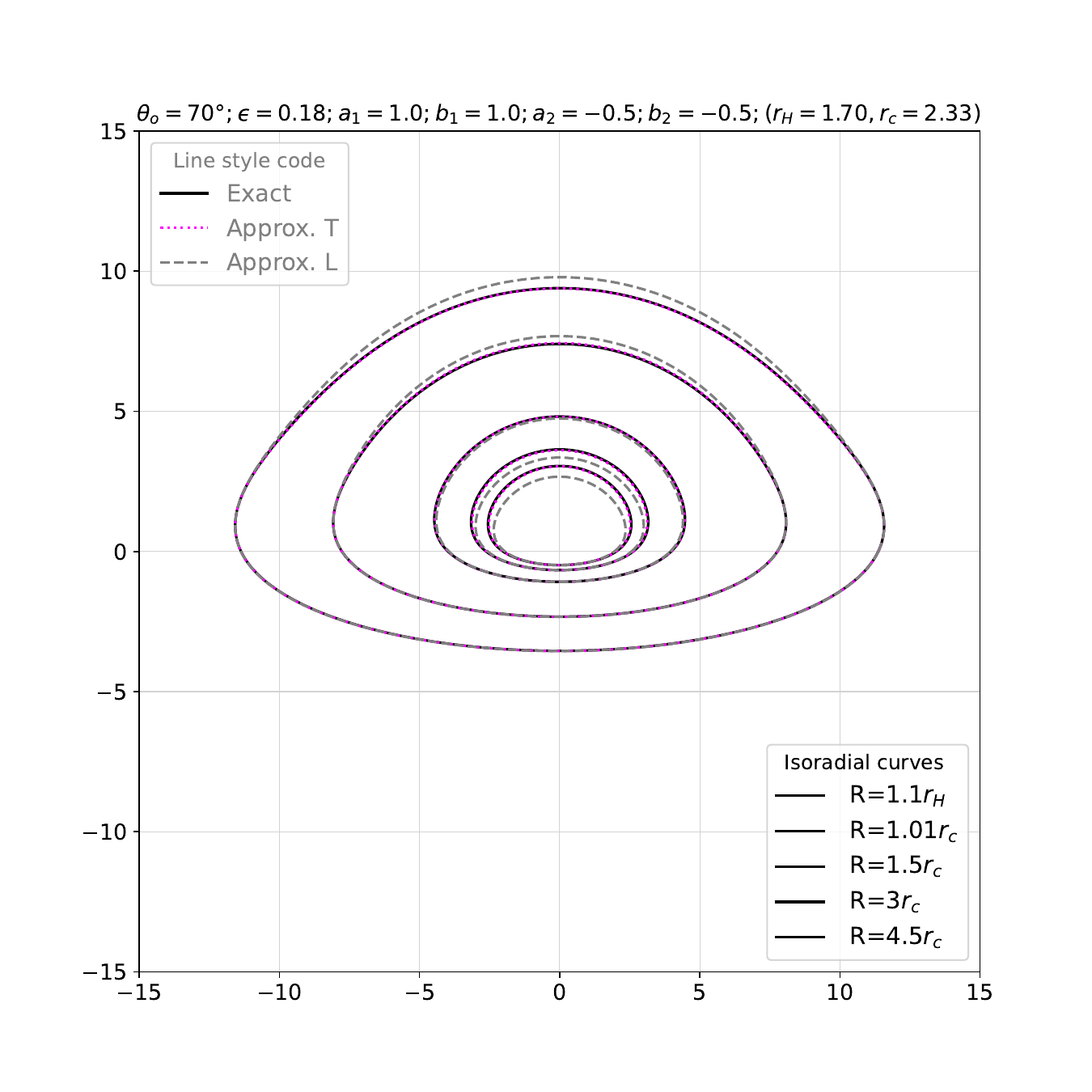}} &
        \hspace{-0.6cm}
        {\includegraphics[scale=0.31,trim={1.7cm 1.65cm 1.26cm 1.47cm},clip]{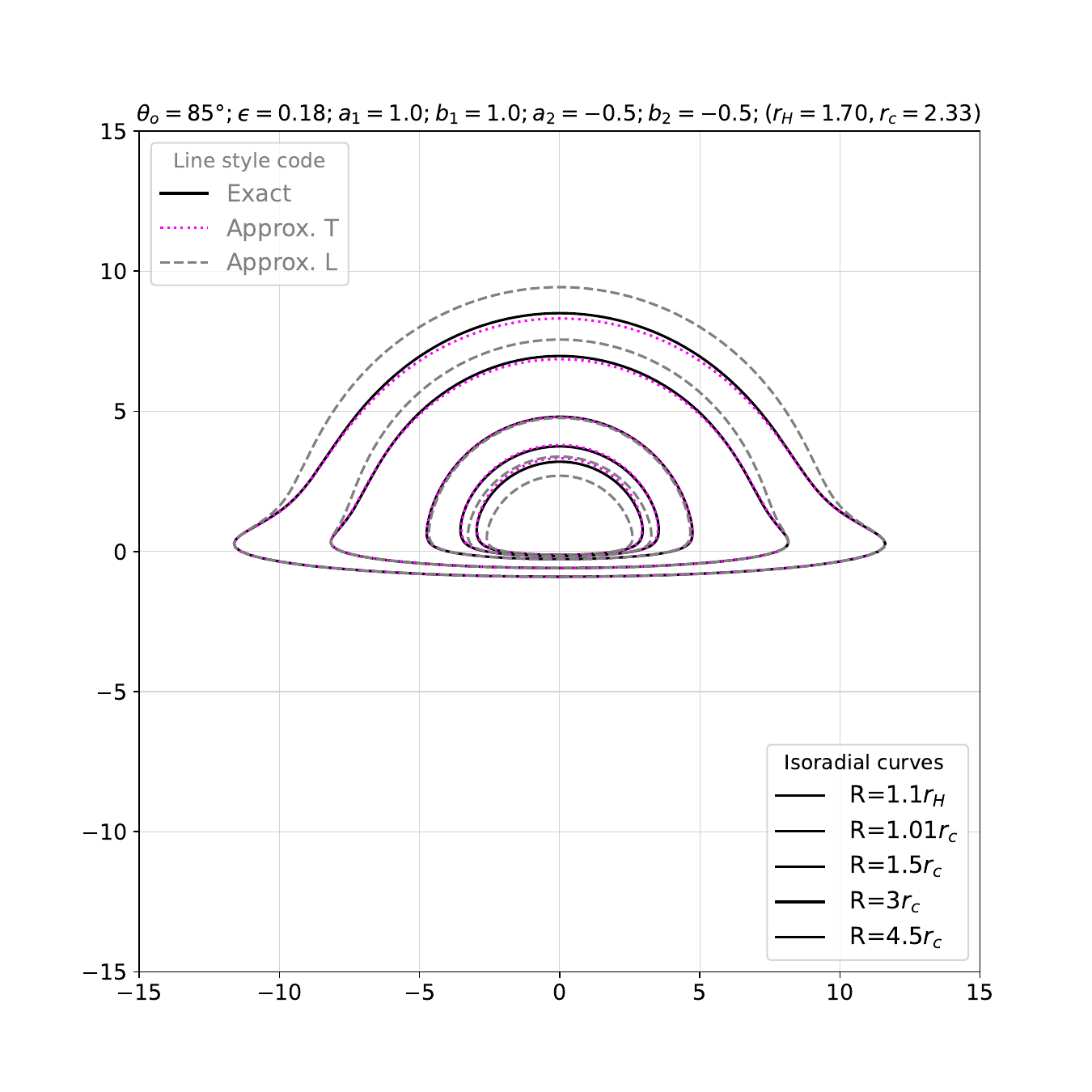}} \\
        \hspace{-9mm}
        {\includegraphics[scale=0.32,trim=1.32cm 1.65cm 1.24cm 1.47cm]{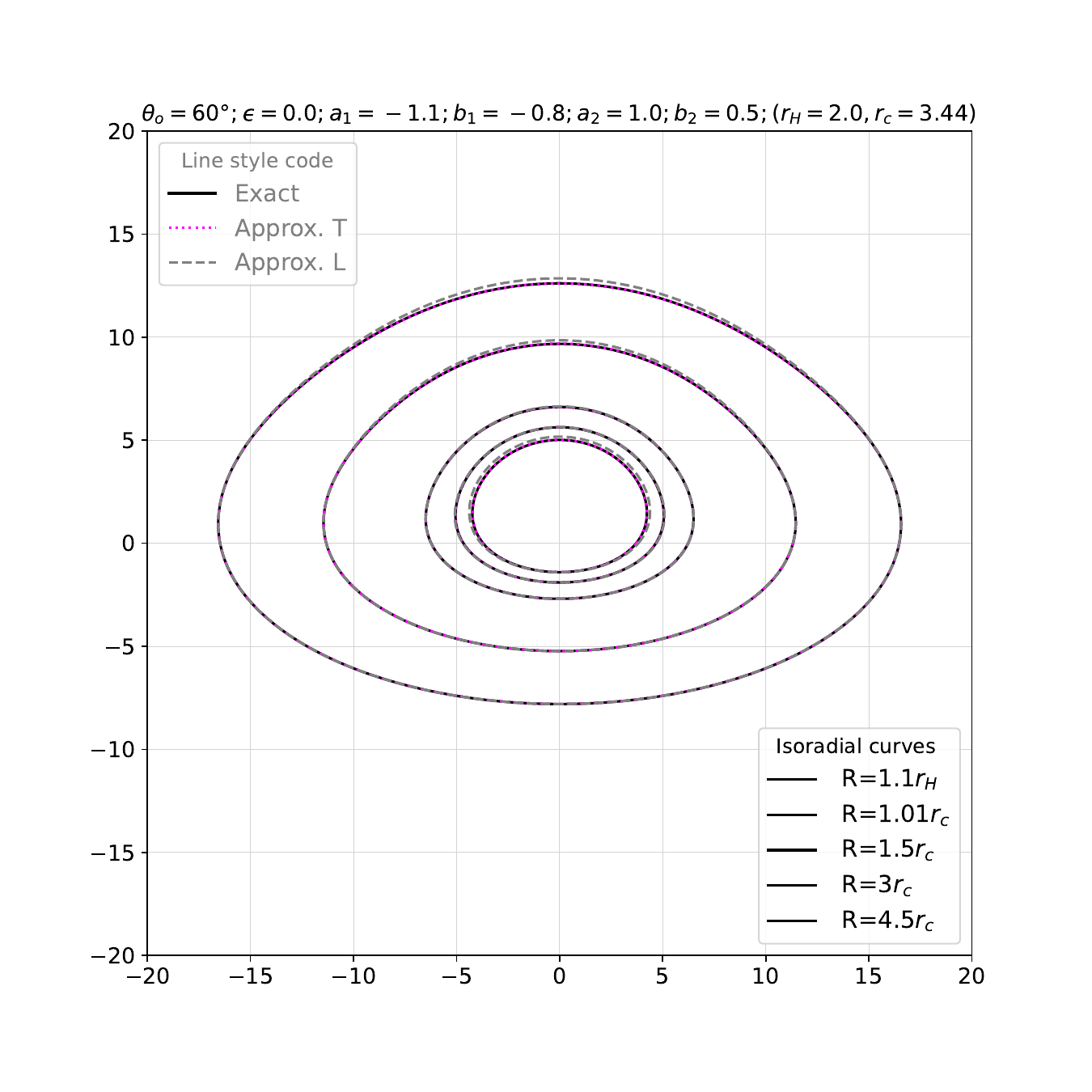}} & 
        \hspace{-0.9cm}
        {\includegraphics[scale=0.32,trim=1.32cm 1.65cm 1.24cm 1.47cm]{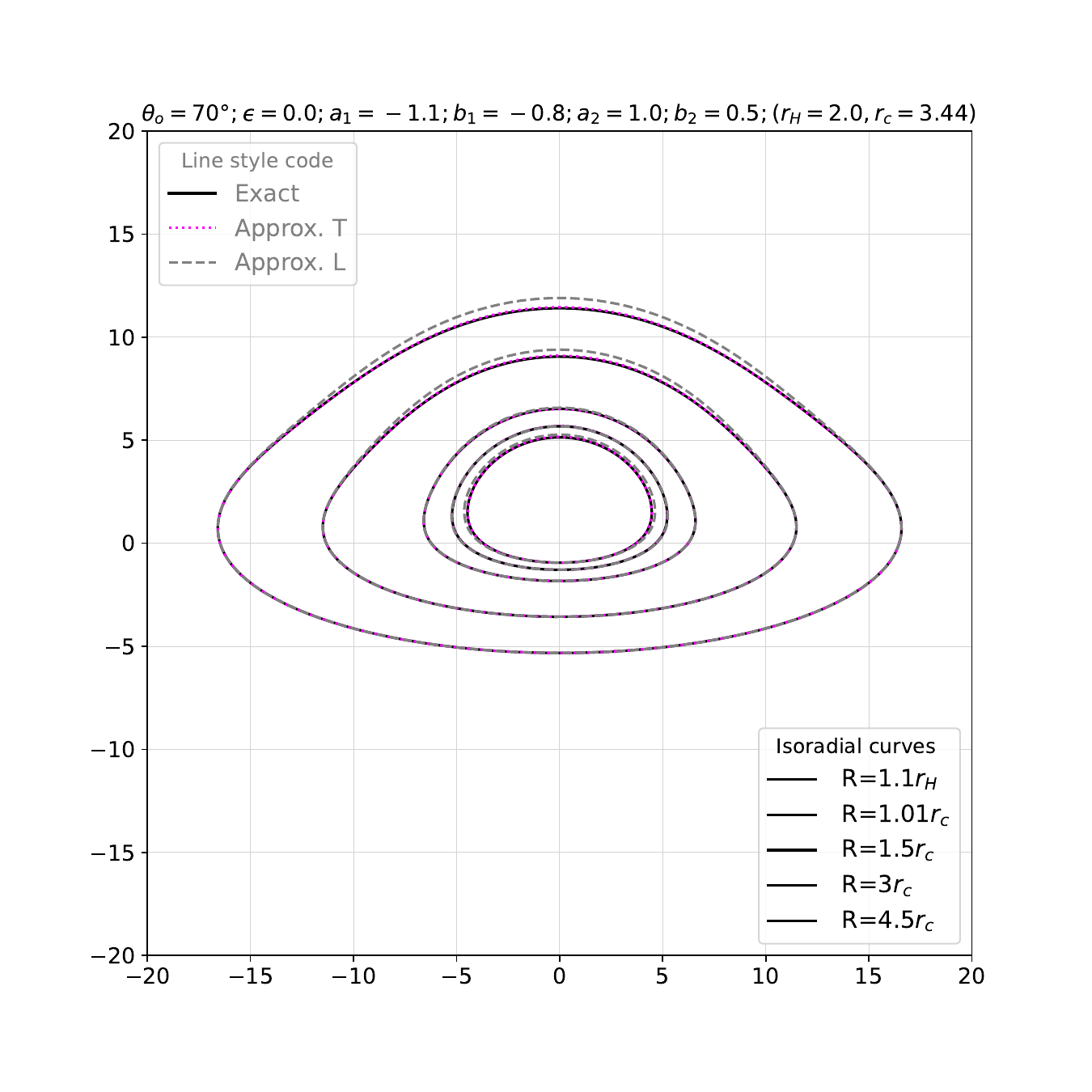}} &  
        \hspace{-0.7cm}
        {\includegraphics[scale=0.32,trim=1.32cm 1.65cm 1.24cm 1.47cm]{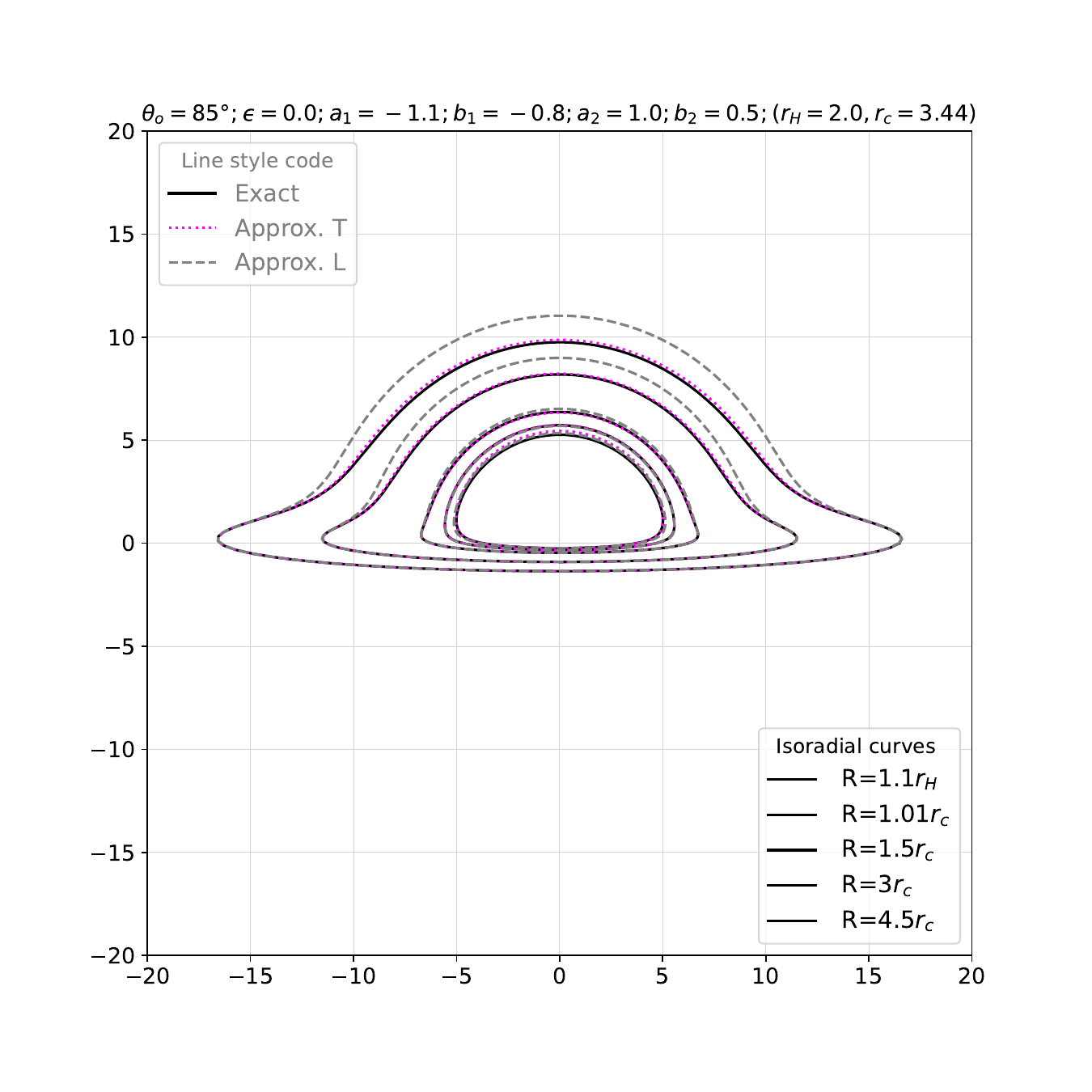}}
    \end{tabular}
    \caption{Similar to Figure \ref{fig:johan_isorad}, this figure depicts a comparison of isoradial curves for the Rezzolla-Zhidenko metric at three distinct disk inclinations. The parameter sets used are: $a_1=1.0; b_1=1.0; a_2=-0.5; b_2=-0.5$ (top row) and $a_1=-1.1; b_1=-0.8; a_2=1.0; b_2=0.5$ (bottom row).} 
    \label{fig:rezzo_isorad}
\end{figure*}

For the RZ metric, the corresponding isoradial curves are presented in Fig.~\ref{fig:rezzo_isorad}. The same values of $R$ used in the previous section are adopted here for direct comparison. The figure illustrates how the image morphology evolves with increasing disk inclination and with variations in the metric parameters $(a_1, a_2, b_1, b_2)$.
 As the inclination angle $\theta_o$ increases, the 
 `Approx.~L'~\eqref{eq:our_aprox_belo} begins to lose accuracy in reproducing the far-side region of the accretion disk, particularly for $\psi \rightarrow \pi$, where light rays travel close to the event horizon before reaching the observer. This deviation is most noticeable in the first row of Fig.~\ref{fig:rezzo_isorad}, corresponding to the case $r_H=1.70$. Here, the simplified linear approximation slightly underestimates the bending of light near the black hole horizon, producing small shape distortions along the far-side contours of the disk image. In contrast, for the outer isoradial curves, the linear approximation tends to overestimate the deflection, leading to marginally larger apparent displacements in the image plane.
  In the case of $r_H=2.0$ and the given choice of parameters $a_1,a_2,b_1,b_2$, with inclinations such that $\theta_o<85^{\circ}$, the approximation of Eq.~\eqref{eq:our_aprox_belo} is indistinguishable from that obtained numerically up to radial coordinate values $r$ less than $1.5r_c$. By contrast, the `Approx.~T'~\eqref{eq:our_aprox_full} retains a high degree of accuracy throughout the entire range of $R$ and $\theta_o$ values explored. Even for the extreme inclination $\theta_o=85^{\circ}$ and the largest emission radii $r\geq3r_c$, the analytical curves remain virtually indistinguishable from the numerical reference, except for a very small deviation at $\psi\rightarrow\pi$. Therefore, the inclusion of higher-order terms in the approximation not only improves the local accuracy of the $\alpha$--$\psi$ relation but also ensures excellent global consistency in the image reconstruction, capturing with a very accuracy the characteristic lensing asymmetry between the near and far sides of the accretion disk.

\begin{figure*}[htbp]
    \centering
    \begin{tabular}{ccc}
        \hspace{-6mm}
        {\includegraphics[scale=0.32,trim=1.5cm 1.2cm 1.5cm 1.2cm,clip]{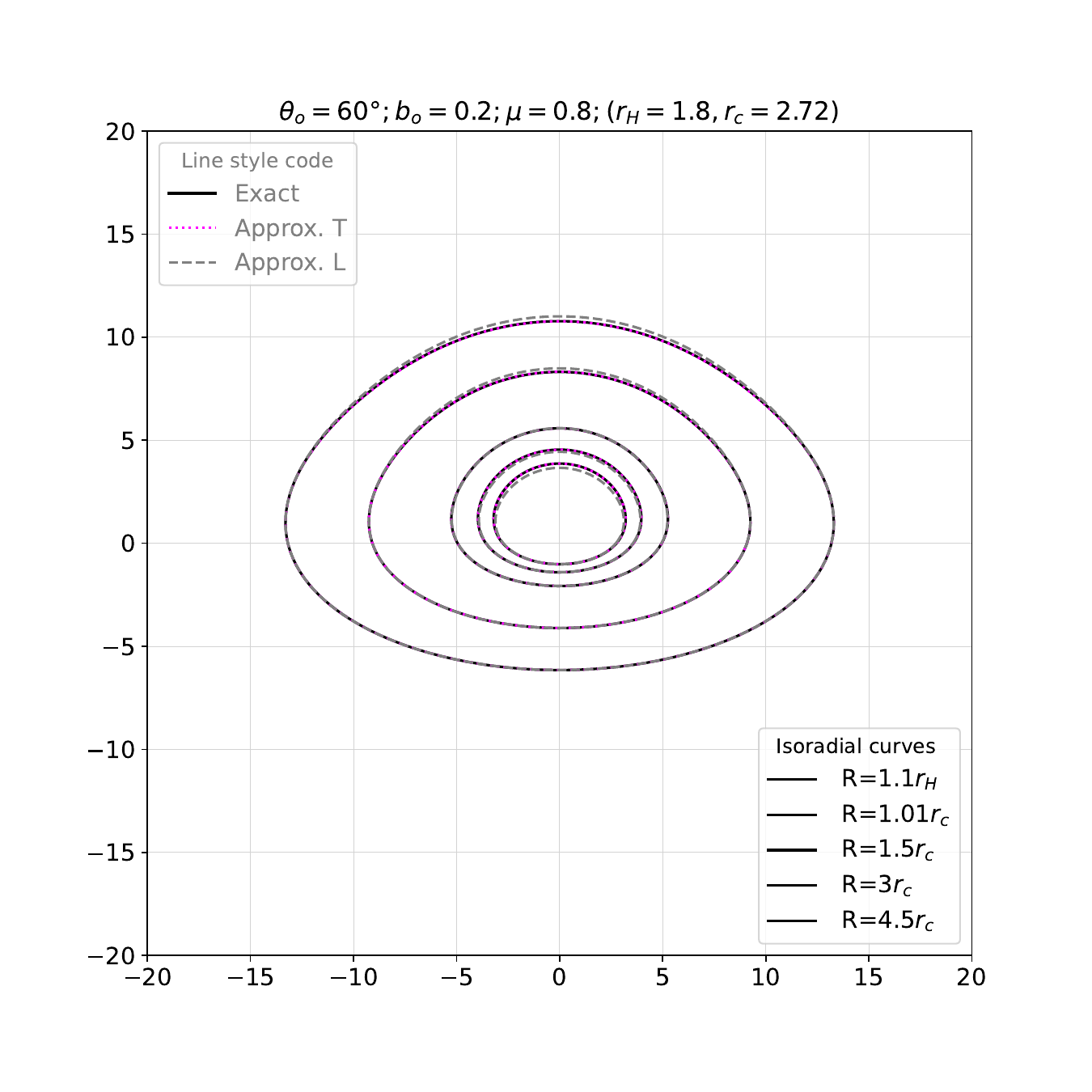}} &
        \hspace{-0.4cm}
        {\includegraphics[scale=0.32,trim=1.5cm 1.2cm 1.5cm 1.2cm,clip]{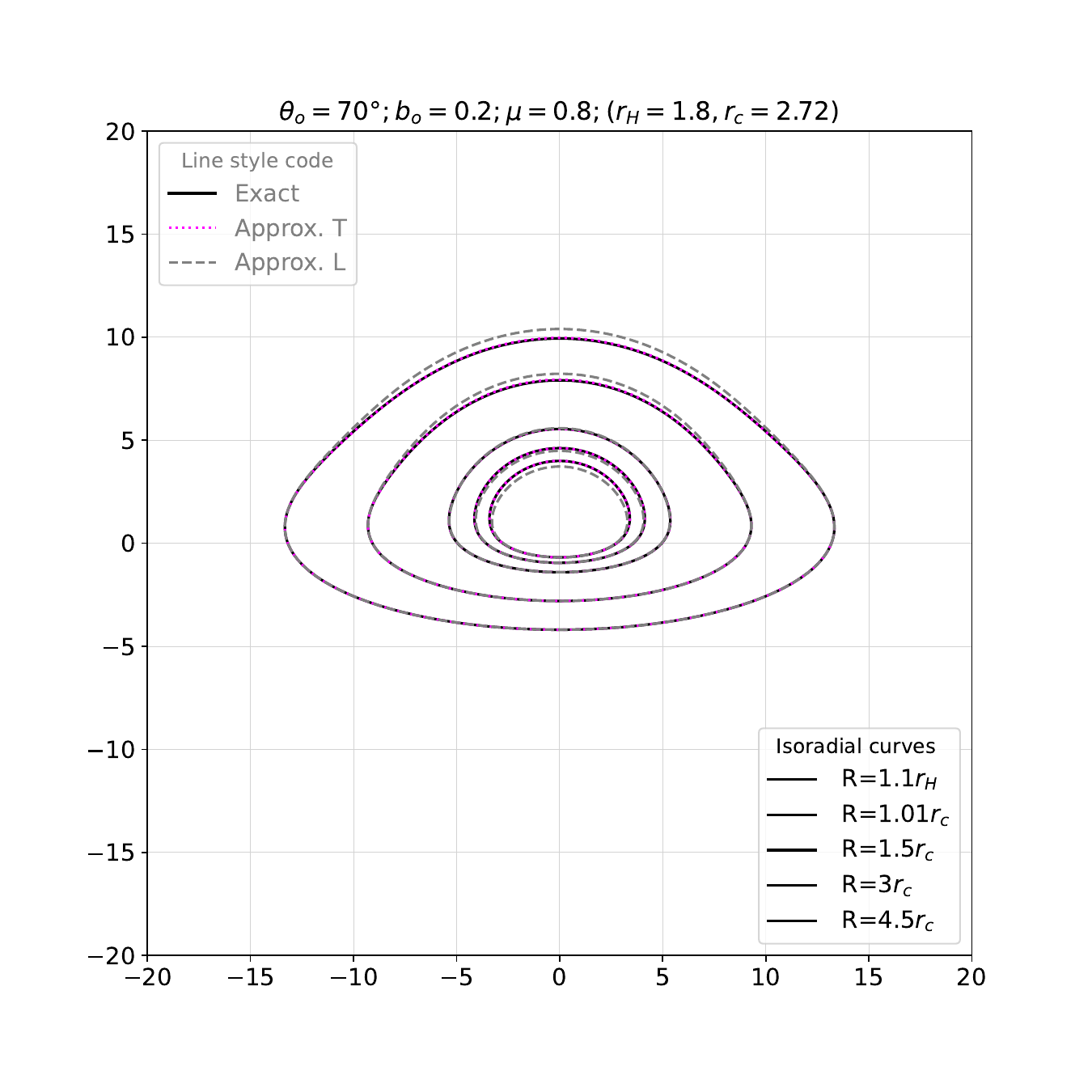}} &
        \hspace{-0.4cm}
        {\includegraphics[scale=0.32,trim=1.5cm 1.2cm 1.5cm 1.2cm,clip]{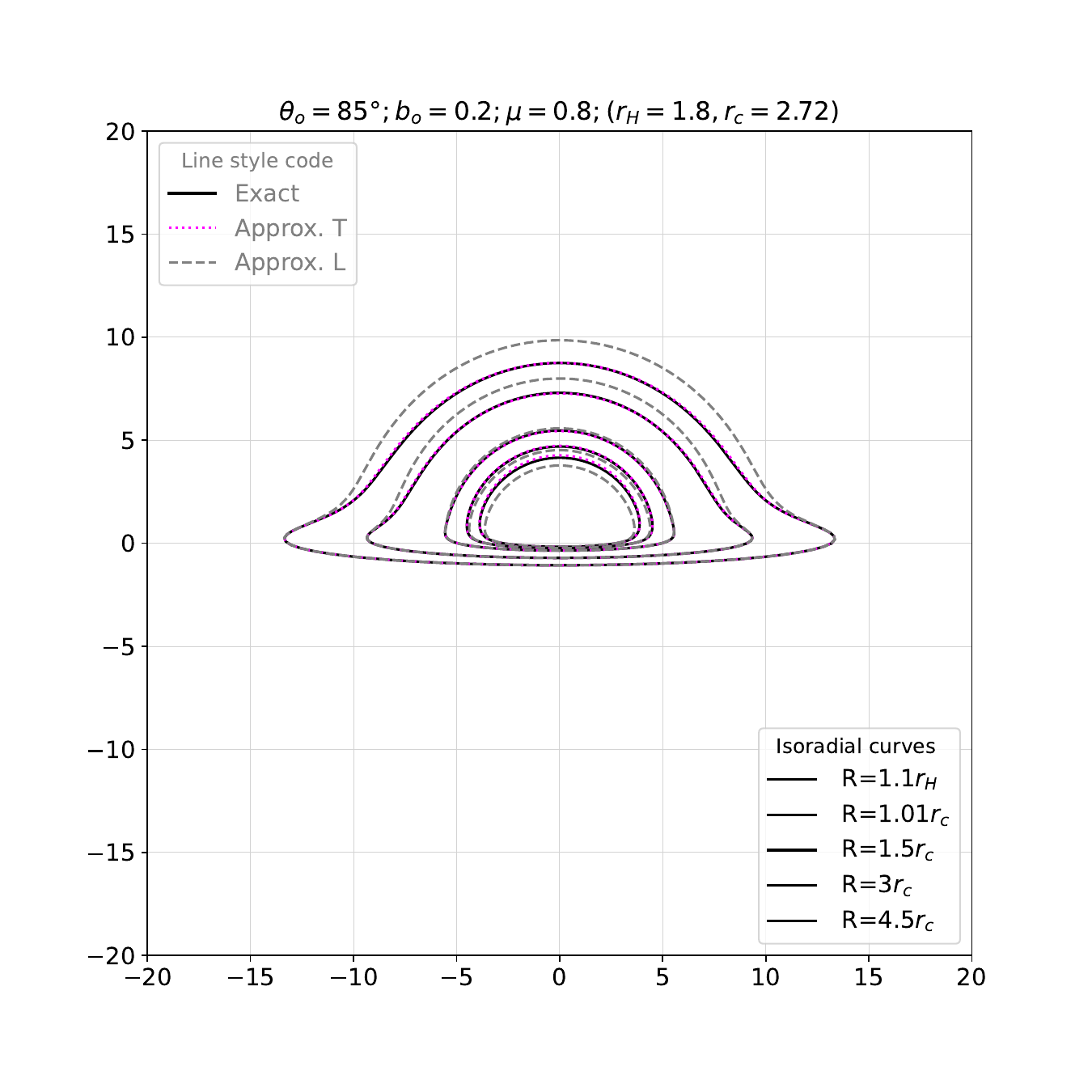}} \\
        \hspace{-6mm}
        {\includegraphics[scale=0.32,trim=1.5cm 1.2cm 1.5cm 1.2cm,clip]{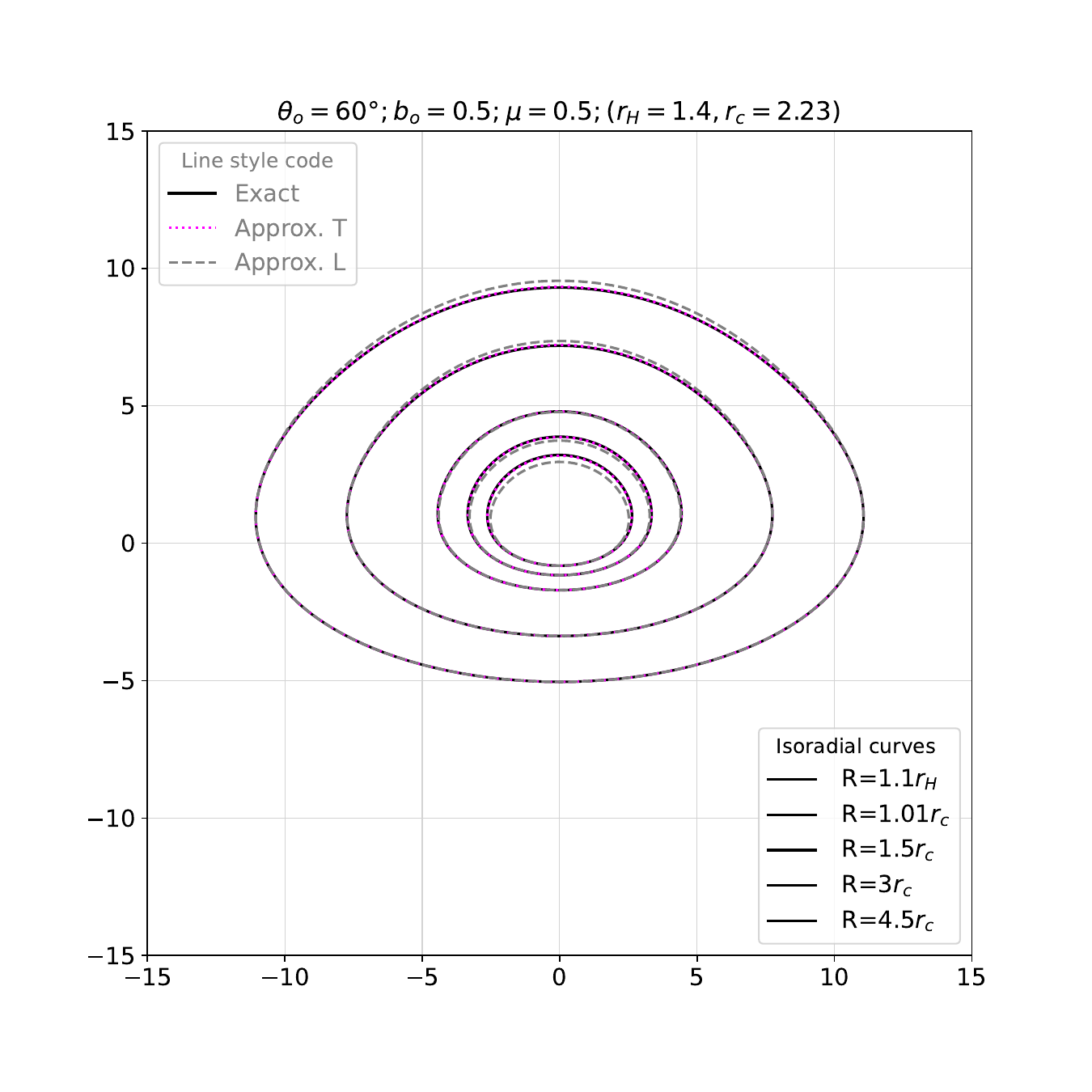}} & 
        \hspace{-0.4cm}
        {\includegraphics[scale=0.32,trim=1.5cm 1.2cm 1.5cm 1.2cm,clip]{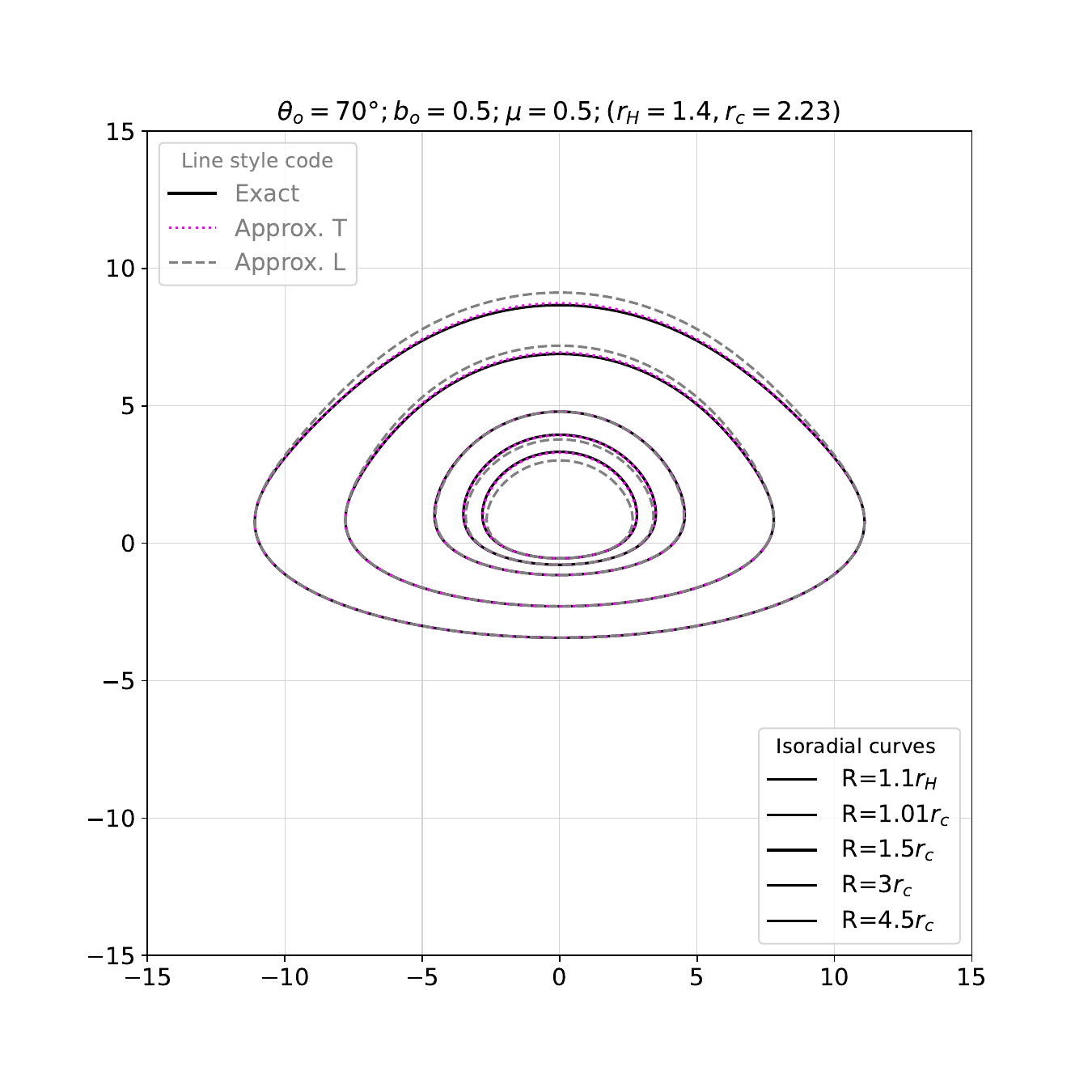}} &  
        \hspace{-0.4cm}
        {\includegraphics[scale=0.32,trim=1.5cm 1.2cm 1.5cm 1.2cm,clip]{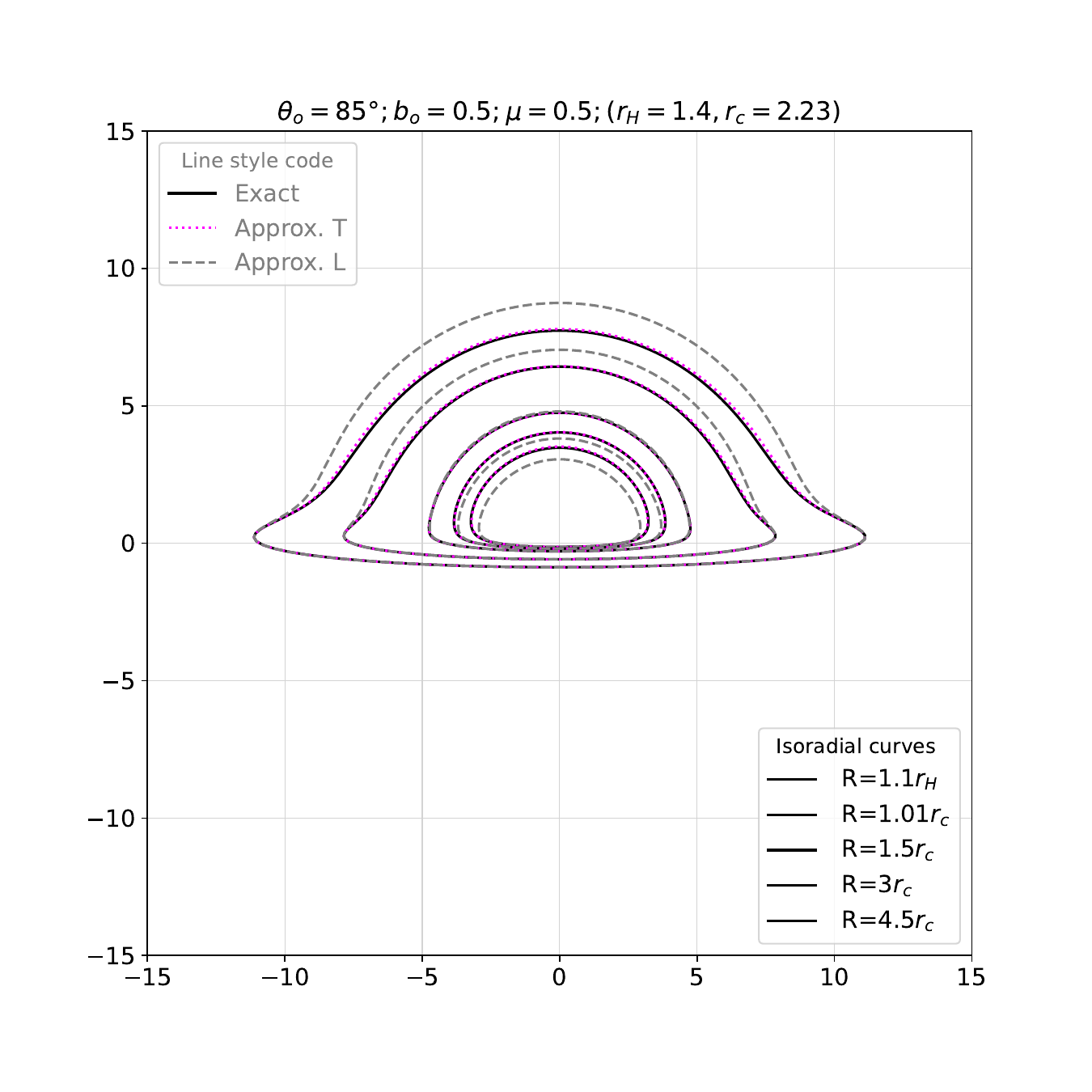}}
    \end{tabular}
    \caption{Similar to Figure \ref{fig:johan_isorad} and \ref{fig:rezzo_isorad}, this plot provides the comparison for Einstein-Maxwell-dilaton-axion metric.}
    \label{fig:emda_isorad}
\end{figure*}

Finally, in the case of the particular EMDA-metric, images of the isoradial curves are shown for two arbitrary sets of parameters: the first corresponding to $b_{o}=0.2$ and $\mu=0.8$, and the second to $b_{o}=0.5$ and $\mu=0.5$, for both sets, $M=1$. The curves are plotted for the same values of $R$ as in the previous cases. It is observed that as the disk inclination increases, the `Approx.~L'~\eqref{eq:our_aprox_belo} loses accuracy. This discrepancy is most evident in the plots with parameters $b_{o}=0.5$ and $\mu=0.5$ , particularly for high inclination angles such as $\theta_o=85^{\circ}$, as shown in bottom row of Fig. \ref{fig:emda_isorad}. In contrast, `Approx.~T'~\eqref{eq:our_aprox_full} maintains high accuracy for all values of $R$ and for all inclination angles analyzed. It is, in fact, practically indistinguishable from the numerically calculated ``exact'' curve, except at an angle of $\theta_o=85^{\circ}$ and for a value very near the event horizon $r = 1.1r_H$, where a slight difference can be observed.

Overall, these comparisons show that the analytical framework developed here reproduces the exact ray-tracing results with remarkable precision for a wide range of parameter values and viewing geometries. This validates the applicability of our approximations to realistic modeling scenarios, such as accretion disk imaging in generic spherically symmetric spacetimes.

\subsection{$QU$-diagrams of orbiting hotspots} \label{sec:qu_diagrams}

As a physical application of the new analytical expressions, Eqs.~\eqref{eq:our_aprox_full} and~\eqref{eq:our_aprox_belo}, we analyze the linear polarization of synchrotron emission produced in the equatorial plane of a compact object and observed at infinity. Following Refs.~\cite{EventHorizonTelescope:2021btj,Claros:2024atw}, we adopt the geometry depicted in Fig.~\ref{fig:gralframe}, focusing on radiation emitted from different points on the equatorial plane and detected by a distant observer.
Within the geometric--optics approximation, and restricting attention to direct rays, the $\alpha$--$\psi$ relations provide a fully analytical mapping from emission radius and direction to the observer’s screen, which we use to compute  Stokes parameters $(Q,U)$ along isoradial curves without resorting to full numerical ray tracing.

We introduce a local static frame centered at point $P$, referred to as the {$P$-frame}, which is associated with the rotational dynamics of the fluid element. The $P$-frame is defined by the tetrad $\{\vb*{\hat{t}}, \vb*{\hat{x}}, \vb*{\hat{y}}, \vb*{\hat{z}}\}$, whose spatial axes are specified as follows: $\vb*{\hat{t}}$ is the unit vector proporcional to the Killing vector $(\frac{\partial}{\partial t})^a$; 
$\vb*{\hat{x}}$ is aligned with $\overline{OP}$; 
$\vb*{\hat{y}}$ is tangent to the circular orbit of radius $R$ at $P$ (i.e., the azimuthal direction); 
and $\vb*{\hat{z}}$ is orthogonal to the disk plane, aligned with the global $Z$-axis.

We next consider a light ray confined to the $\overline{POO'}$ plane, shown as the blue plane in the inset of Fig.~\ref{fig:gralframe}. 
This plane is described by the coordinate system $(x',y',z')$, where the $x'$-axis is oriented along $\overline{OP}$, and the observer $O'$ lies in the $x'z'$-plane. 
These spatial coordinates are complemented by a temporal coordinate  $t$. 
The set $(t,x',y',z')$, together with its associated orthonormal tetrad $\{\vb*{\hat{t}}, \vb*{\hat{x}'}, \vb*{\hat{y}'}, \vb*{\hat{z}'}\}$ defined at point $P$, specifies a locally static {$G$-frame}. 
In the $G$-frame, the photon's four-momentum $\vb*{k}_{(G)}$ forms an emission angle $\alpha$ with respect to the $\vb*{\hat{x}'}$-axis. 
Since $\vb*{k}_{(G)}$ lies entirely within the $x'z'$-plane, its projection along the $\vb*{\hat{y}'}$-axis vanishes~\cite{Claros:2024atw}. 
The relation between the $P$- and $G$-frames is given by a spatial rotation about the $\vb*{\hat{x}}$-axis through an angle $\xi$ (see Eq.~(5.29) of Ref.~\cite{Claros:2024atw}).

Finally, we introduce the fluid $F$-frame, which is co-moving with the fluid element orbiting the black hole. This frame is defined by the orthonormal tetrad $\{\vb*{\tilde{t}}, \vb*{\tilde{x}}, \vb*{\tilde{y}}, \vb*{\tilde{z}}\}$ and is related to the $P$-frame through a Lorentz transformation. Further details regarding the relationships among the aforementioned frames and the expression of the four-null vector $\vb*{k}_{(F)}$ in the $F$-frame can be found in Ref.~\cite{Claros:2024atw}.

In order to compute the synchrotron radiation emitted by the fluid element in the $F$-frame, it is necessary to determine the polarization $4$-vector $\vb*{f}$. For this purpose, the magnetic field $\vb*{B}$ in the fluid frame is required, and is given by

\begin{equation} \label{eq:field_b}
    \begin{aligned}
        &\vb*{B}=B_{r}\vb*{\tilde{x}} +B_{\phi}\vb*{\tilde{y}}+B_{z}\vb*{\tilde{z}}=\vb*{{B}}_{eq}+B_{z}\vb*{\tilde{z}}, \\
        &\vb*{{B}}_{eq}:= B_{r}\vb*{\tilde{x}} +B_{\phi}\vb*{\tilde{y}}=B_{eq}(\cos\eta \ \vb*{\tilde{x}} + \sin\eta \ \vb*{\tilde{y}}), \nonumber
    \end{aligned}  
\end{equation}
where $\eta$ is measured with respect to $\vb*{\tilde{x}}$. Here we set the norm of $\vb{B}$ to 1. Therefore, $\vb*{f}$ can be expressed in the $F$-frame as 

\begin{equation} \label{eq:fframe_pol}
    \begin{aligned}
        &f^{\hat{\tilde{t}}}_{(F)}=0, \\       &f^{i}_{(F)}=\frac{(\vb*{k}_{(F)}\times\vb*{B})^{i}}{|\vb*{k}_{(F)}|}, \hspace{4mm} i=\hat{\tilde{x}},\hat{\tilde{y}},\hat{\tilde{z}}.
    \end{aligned}  
\end{equation}
where $\vb*{k}_{(F)}$ represent the spatial projection of photon's momentum in the $F$-frame. Once the polarization four-vector $\vb*{f}$ has been computed in the $F$-frame, the tetrad components of both $\vb*{f}$ and $\vb*{k}$ can be obtained using the metric~\eqref{eq:metric_spheric}~\cite{EventHorizonTelescope:2021btj,Claros:2024atw}. To transform back to the $P$-frame, the inverse Lorentz transformations are applied.

The velocity of the fluid element, $\vb*{\beta}$, in the $P$-frame can be written as
\vspace{-2mm}
\begin{equation} \label{eq:beta}
\vb*{\beta} = \beta(\cos\chi \vb*{\hat{x}} + \sin\chi \vb*{\hat{y}}),
\end{equation}
where the angle $\chi$ is measured with respect to the unit vector $\vb*{\hat{x}}$, and the motion lies within the disk plane. As is well known~\cite{Walker:1970un,Frolov:2017kze}, along an affinely parametrized null geodesic, as the polarization vector $\vb*{f}$ is both orthogonal and parallel transported, the following quantities remain constant along the geodesic:
\begin{equation} \label{eq:walkpenrose_const_or}
    \begin{aligned}
        &\kappa = \kappa_{1}+i\kappa_{2}, \\
        &\kappa_{1} = \sqrt{A(r)B(r)}r(k^{t}f^{r}-k^{r}f^{t}), \\ 
        &\kappa_{2} = r^{3}\sin\theta(k^{\theta}f^{\phi}-k^{\phi}f^{\theta}),
    \end{aligned}  
\end{equation}
which, when evaluated at point $P$ and expressed in terms of the tetrad components of $\vb*{k}$ and $\vb*{f}$ in the $P$-frame, reduce to the following expressions:
\begin{equation} \label{eq:walkpenrose_const}
    \begin{aligned}
        &\kappa_{1}=R\left(k^{\hat{t}}_{(P)}f^{\hat{x}}_{(P)}-k^{\hat{x}}_{(P)}f^{\hat{t}}_{(P)}\right), \\ &\kappa_{2}=R\left(k^{\hat{y}}_{(P)}f^{\hat{z}}_{(P)}-k^{\hat{z}}_{(P)}f^{\hat{y}}_{(P)}\right),
    \end{aligned}  
\end{equation}
with the tetrad components of $\vb*{f}$ and $\vb*{k}$ given in equations $(5.30)$ and $(5.39)$ of \cite{Claros:2024atw}.
   
To obtain the polarization in the asymptotic observer's frame, we use the conserved quantities in Eq.~\eqref{eq:walkpenrose_const} to express the observed electric field in the $X'Y'$ frame as proportional to the vector $(E_{X'},E_{Y'})$ in the image plane
given by
\begin{equation} \label{eq:efield}
    \begin{aligned}
        E_{X'}&=\frac{Y'\kappa_{2}+X'\kappa_{1}}{X'^{2}+Y'^{2}}, \\
        E_{Y'}&=\frac{Y'\kappa_{1}-X'\kappa_{2}}{X'^{2}+Y'^{2}}. \\
    \end{aligned}  
\end{equation}

\begin{figure*}[htbp]
    \centering
    \begin{tabular}{ccc}
        \hspace{-6mm}
        {\includegraphics[scale=0.3,trim=0 0 0 0,clip]{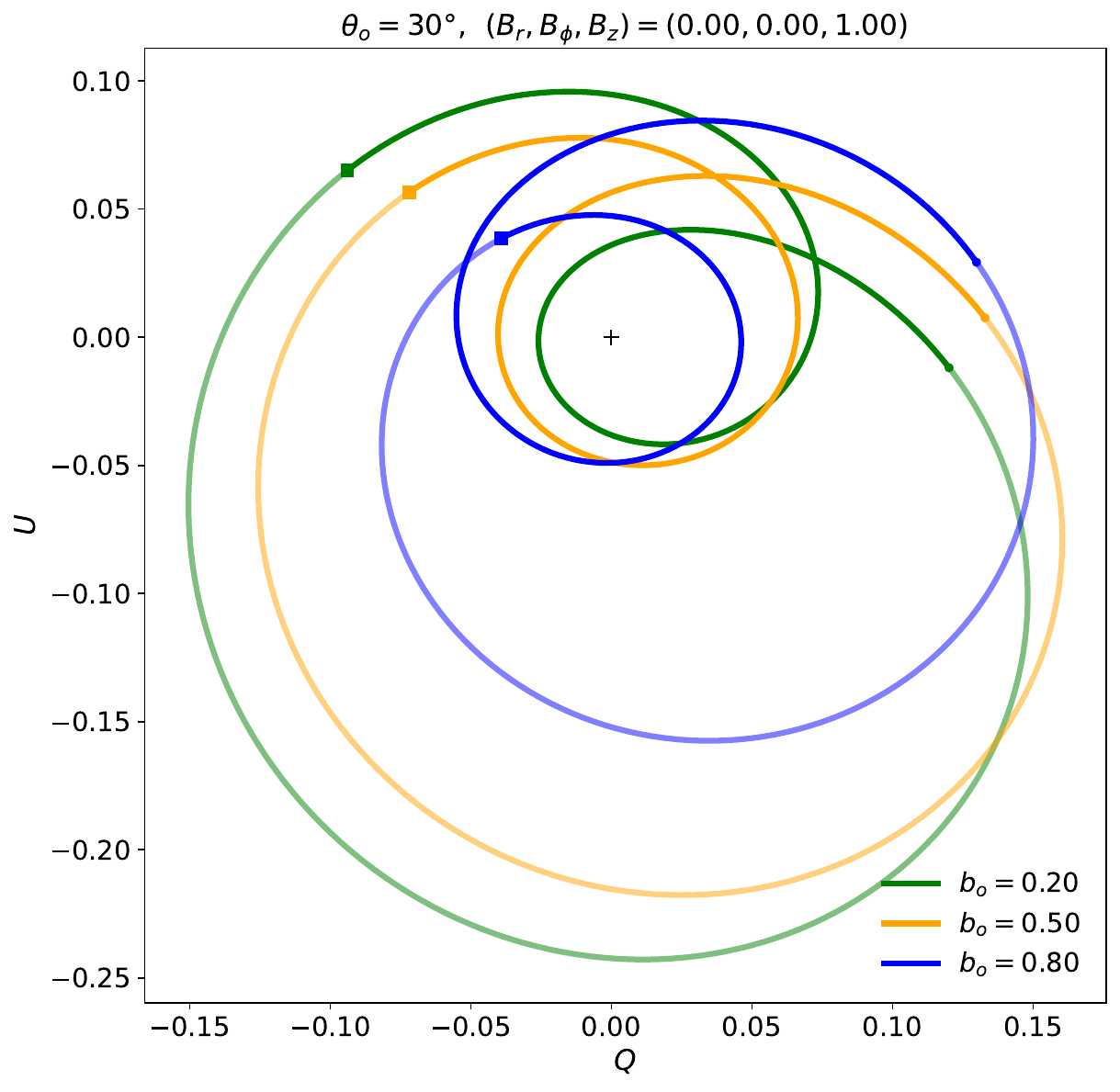}} &
        \hspace{-0.1cm}
        {\includegraphics[scale=0.3,trim=0 0 0 0,clip]{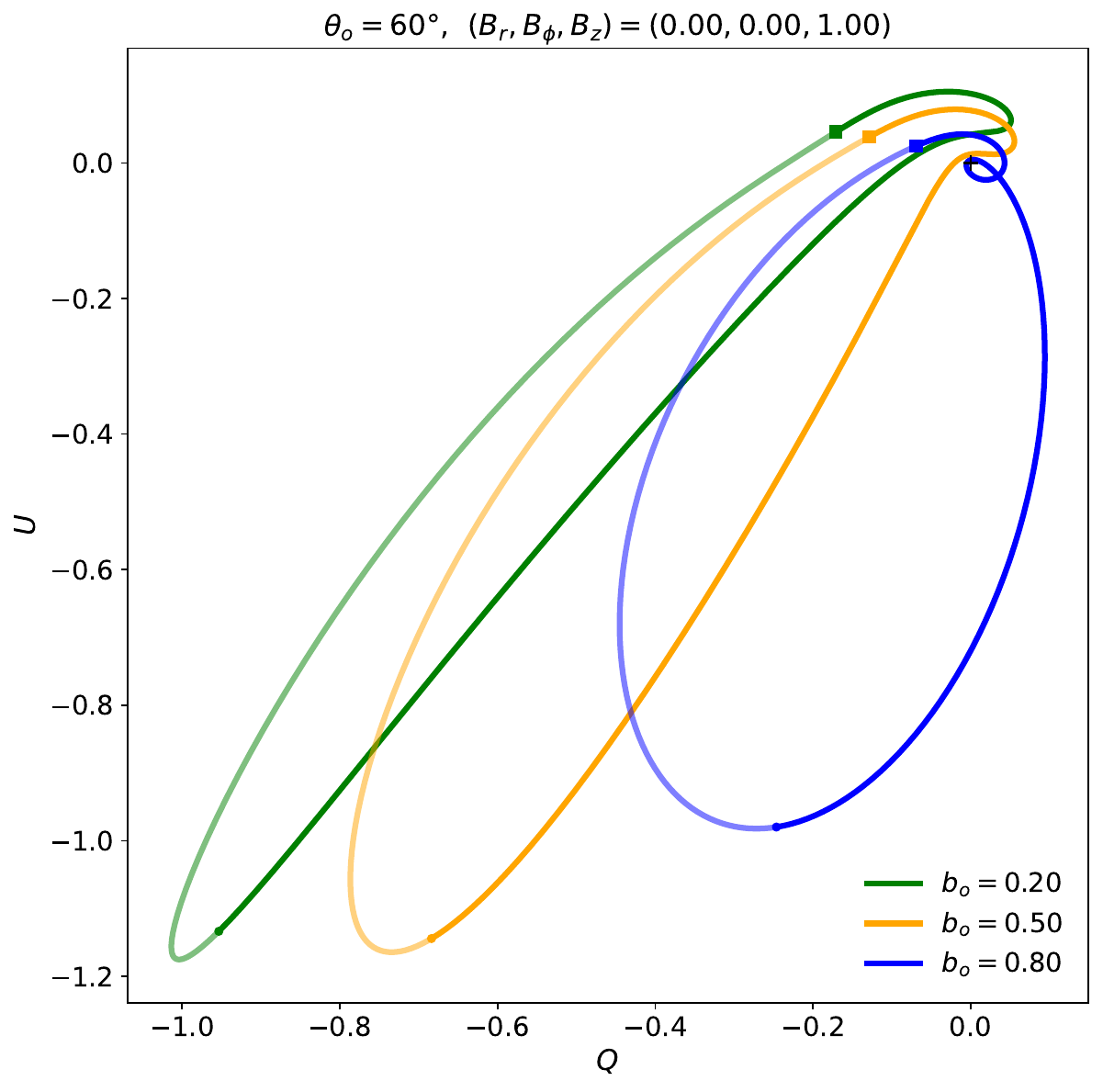}} &
        \hspace{-0.15cm}
        {\includegraphics[scale=0.3,trim=0 0 0 0,clip]{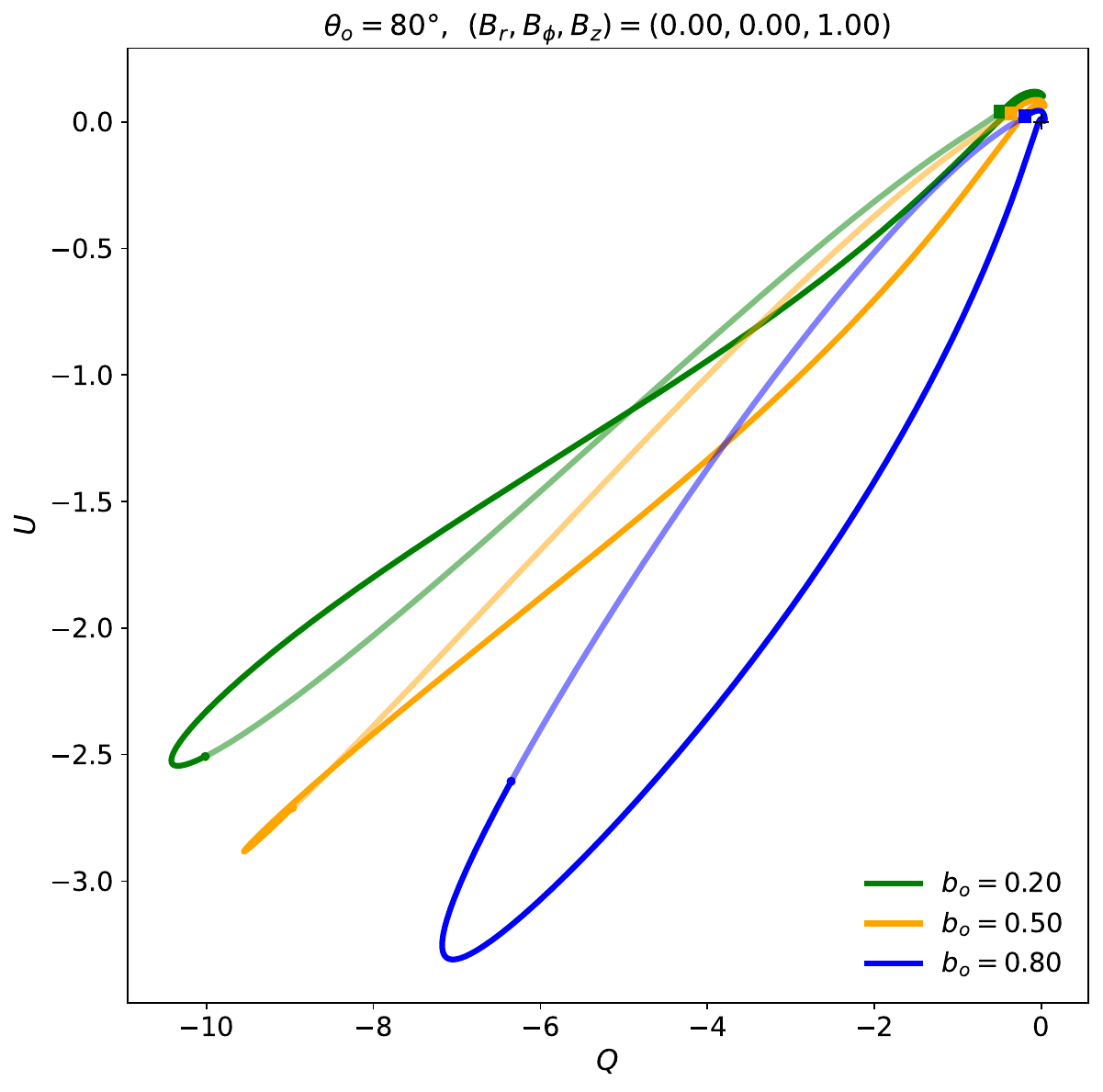}} \\
        \hspace{-6mm}
        {\includegraphics[scale=0.3,trim=0 0 0 0,clip]{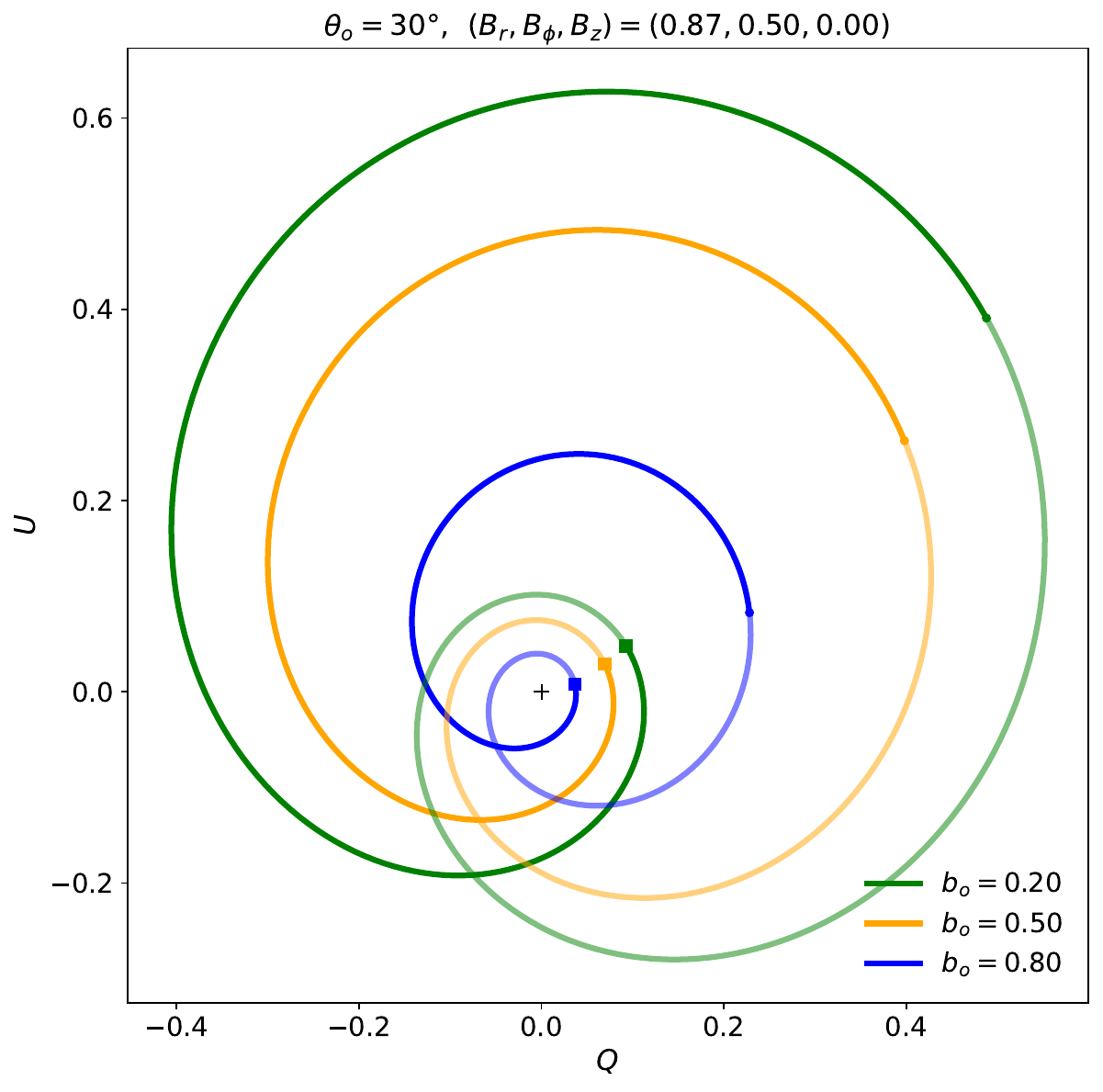}} & 
        \hspace{-0.1cm}
        {\includegraphics[scale=0.3,trim=0 0 0 0,clip]{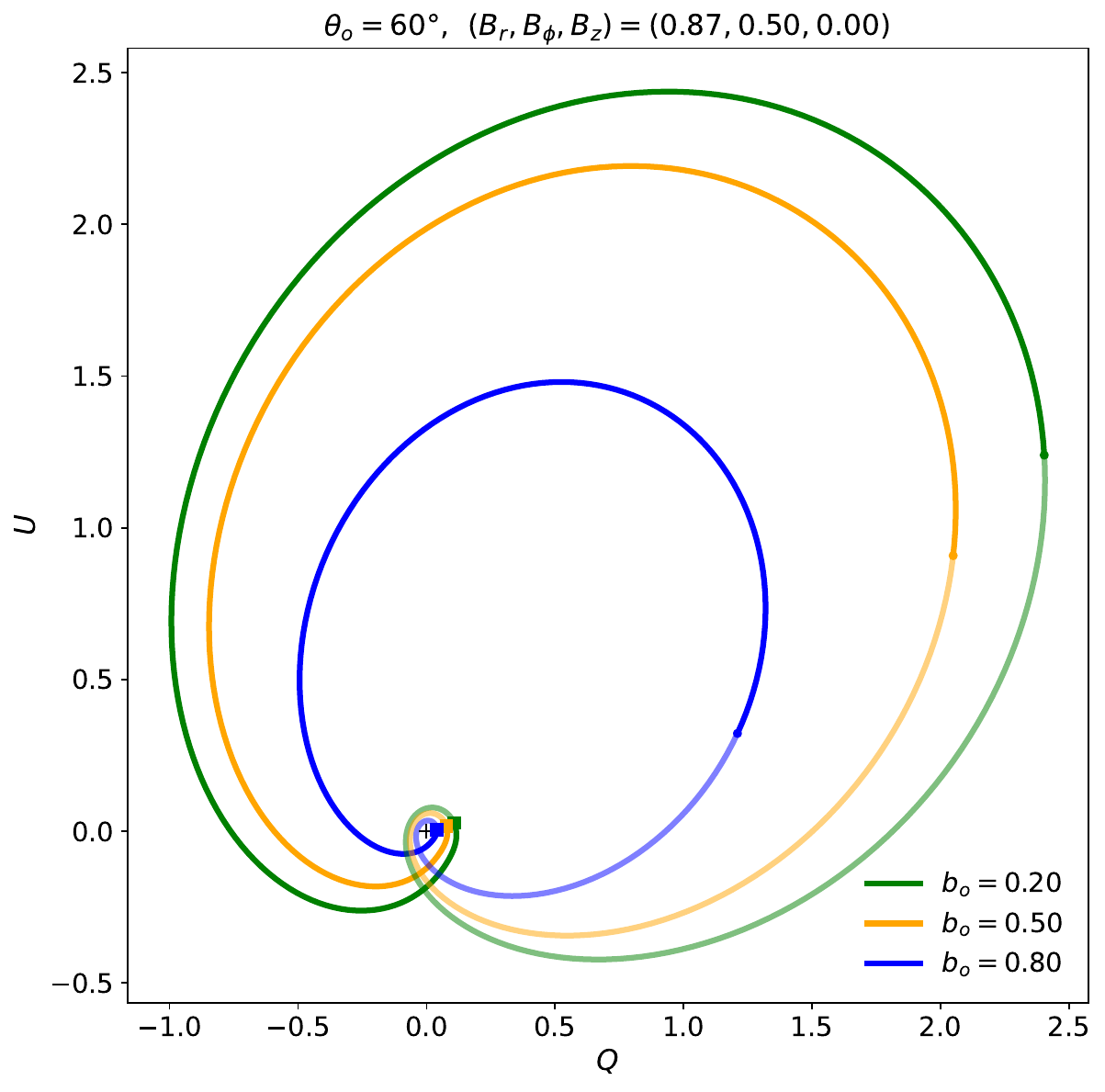}} &  
        \hspace{-0.15cm}
        {\includegraphics[scale=0.3,trim=0 0 0 0,clip]{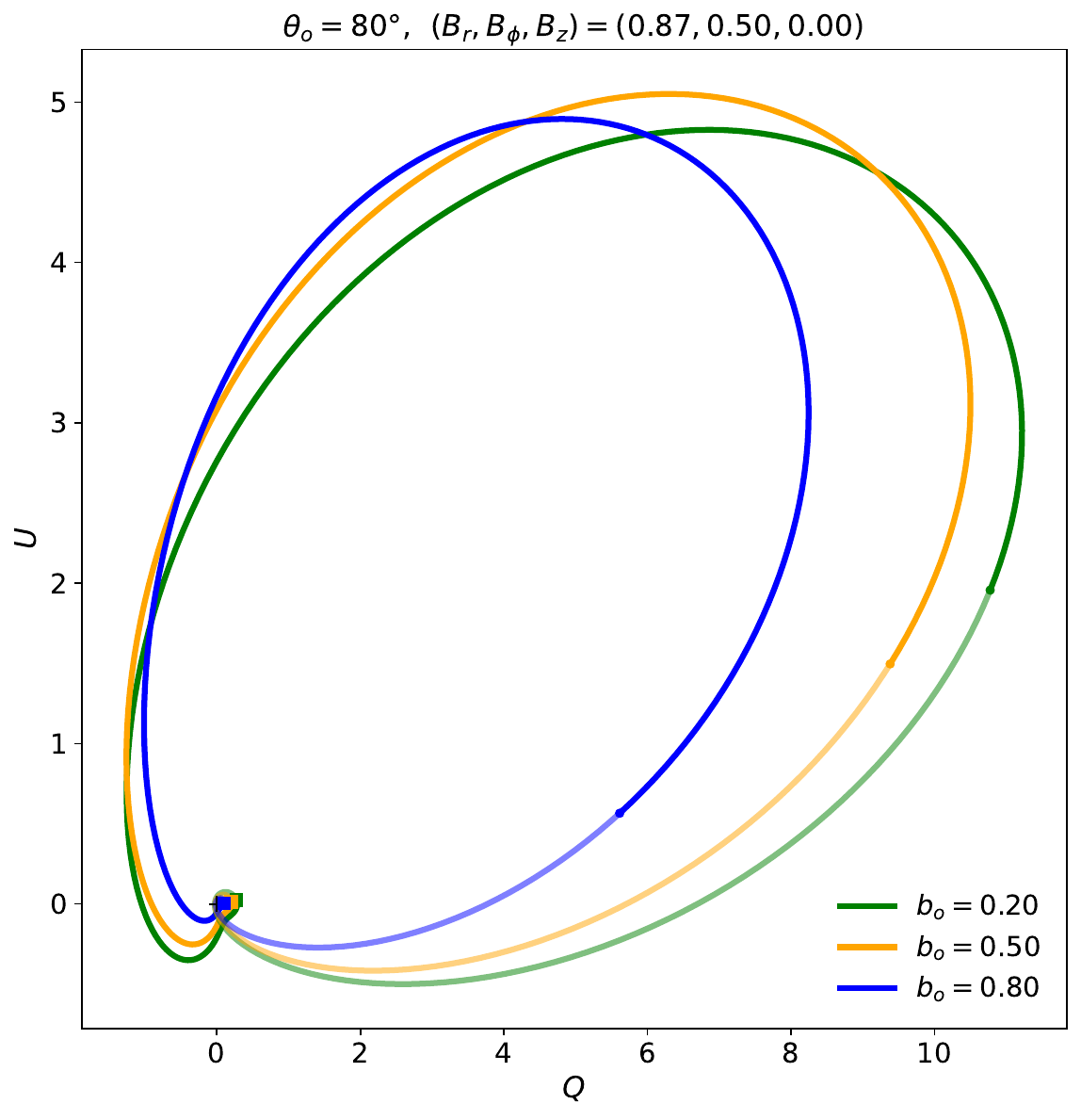}}
    \end{tabular}
    \caption{$QU$-diagrams depict a hotspot, modeled as a point source, orbiting at ISCO of a EMDA black hole for dilaton parameter values $0.20, 0.50$, and $0.80$. Three distinct disk inclinations are considered. The top row illustrates a pure vertical magnetic field $(\vb*{{B}}_{eq}=0)$, while the bottom row represents a pure toroidal field  $(B_{z}=0)$. The hotspot orbits clockwise, starting from colored circle, and the origin $(Q=U=0)$ is denoted by a black crosshair.}    
    \label{fig:qu_emda}
\end{figure*}

Within the framework of an optically and geometrically thin disk model, we consider fluid elements on circular orbits around the black hole that emit linearly polarized synchrotron radiation~\cite{PhysRevLett.28.998,Rybickilightman1979}. Before reaching the observer, the signal is modified by gravitational redshift and Doppler beaming, effects that are codified in the factor
\begin{equation} \label{eq:delta}
\delta = \frac{E_{o}}{E_{(F)}} = \frac{1}{k^{\hat{\tilde{t}}}_{(F)}},
\end{equation}
where $E_{o}$ denotes the photon energy measured by a distant observer, set to $1$, while $E_{(F)}$ represents the photon energy in the $F$-frame. The temporal component $k^{\hat{\tilde{t}}}_{(F)}$ is given in Eq.~(5.32) of Ref.~\cite{Claros:2024atw}. Assuming, in turn, a photon path length $l_p$ within the accretion disk and a spectral index equal to $1$, the observed electric-field components can be expressed as~\cite{EventHorizonTelescope:2021btj,Gelles:2021kti,Claros:2024atw}:

\begin{equation} \label{eq:efield_obs}
    \begin{aligned}
        E_{X',obs}&=\delta^{2}l_{p}^{1/2}E_{X'}, \\
        E_{Y',obs}&=\delta^{2}l_{p}^{1/2}E_{Y'}.
    \end{aligned}  
\end{equation}
Thus, the total polarized intensity and the electric vector position angle (EVPA) are given by
\begin{equation} \label{eq:stokes_1} \begin{aligned} \mathrm{P}&=E_{X',\text{obs}}^{2}+E_{Y',\text{obs}}^{2},\\ \text{EVPA}&=\arctan\frac{E_{Y',\text{obs}}}{E_{X',\text{obs}}}=\frac{1}{2}\arctan\frac{U}{Q}, \end{aligned} \end{equation}
where the Stokes parameters $Q$ and $U$ are defined as
\begin{equation} \label{eq:stokes_2}
    \begin{aligned}
        Q&= E_{X',\text{obs}}^{2}-E_{Y',\text{obs}}^{2}, \\
        U&=2E_{X',\text{obs}}E_{Y',\text{obs}}. \\
    \end{aligned}  
\end{equation}

Note that, according to Eq.~\eqref{eq:stokes_2}, the polarization angle is defined with respect to the $X'$ axis. In astronomical practice, however, it is more common to define it with respect to the $Y'$ axis. Since these conventions differ by a $90^\circ$ rotation of the reference axes, the associated Stokes parameters $(Q,U)$ acquire an overall minus sign when moving from the $X'$-based to the $Y'$-based convention.

Using the formalism described above, we model a hotspot orbiting in the vicinity of a black hole as a fluid element emitting synchrotron radiation just outside the innermost stable circular orbit (ISCO). This setup is consistent with observations of flaring events attributed to hotspots near the ISCO in systems with sufficient temporal resolution. For a spacetime described by the metric in Eq.~\eqref{eq:metric_spheric}, the tangential velocity of a particle on a Keplerian orbit is given by:

\begin{equation} \label{eq:omega_beta}
        \beta(r)=\frac{r\ \Omega(r)}{\sqrt{A(r)}}  , \quad \Omega(r)= \sqrt{\frac{A'(r)}{2r}}.  
\end{equation}
The ISCO's radial coordinate is calculated using Equation $(5.17)$ of \cite{Claros:2024atw}, assuming clockwise motion ($\chi=-\pi/2$ when the disk is viewed face-on). 

While this polarization model can be readily applied to the metric families in Sec.~\ref{sec:metrics_fam}, this work will focus exclusively on the study of $QU$-diagrams for the particular metric \eqref{eq:emda_metric}. 
{ Three representative values of the dilaton parameter $b_0$ are considered: $0.20$, $0.50$, and $0.80$. For a fixed mass, the parameter $\mu$ is uniquely determined. The corresponding ISCO radii are located approximately at $5.376$, $4.318$, and $2.946$, with the corresponding orbital velocities $\beta$ being approximately $0.518$, $0.557$, and $0.631$, respectively.}

{ In all cases, we adopt the standard radio-astronomical convention: the polarization angle is measured from the 
$Y'$-axis in a counterclockwise direction with $Q$ and $U$ defined as minus the expressions in Eqs.~\eqref{eq:stokes_1} and \eqref{eq:stokes_2} (see, for example, Fig.~1 of \cite{Vincent:2023sbw}).} In the diagrams, colored circles mark $\phi=0$ (or $\phi=2\pi$), while squares indicate $\phi=\pi$. The orbital motion is assumed to be clockwise: starting at the circle, following the light solid curve to the square, and returning to the circle along the dark solid curve.

Figure~\ref{fig:qu_emda} displays the $QU$-diagrams for two distinct magnetic field configurations and three different disk inclinations. The top row corresponds to a purely vertical magnetic field, while the bottom row represents a purely equatorial (toroidal) configuration. In both cases, the hotspot is modeled as a localized synchrotron-emitting source orbiting just above the ISCO of an EMDA black hole, for three representative values of the dilaton parameter $b_0 = \{0.20,\, 0.50,\, 0.80\}$.

For the purely vertical field ($\vb*{B}_{eq}=0$), the polarization tracks exhibit two well-defined loops in the $QU$-plane at low inclination angles. As the inclination increases, the inner loop gradually contracts and eventually merges with the origin, while the outer loop becomes increasingly elongated, evolving into a distorted lemniscate that in most cases does not enclose the origin (except for $b_0=0.80$). This behavior reflects a reduction in the total rotation of the polarization vector during one orbital period, consistent with the EVPA completing less than one full rotation and reversing its sense of rotation once per orbit~\cite{Gelles:2021kti}. 

In contrast, for the equatorial (toroidal) magnetic field configuration ($B_{z}=0$), both loops persist for all disk inclinations. In this case, as the viewing angle increases, the loops expand and separate rather than collapsing, showing a more complex modulation of the EVPA. The polarization vector in this configuration rotates twice during each full orbit, producing the characteristic double-loop pattern previously noted in models with azimuthal magnetic dominance. The larger separation between the loops at higher inclinations highlights the stronger contribution of Doppler boosting to the observed polarization state.

For a fixed inclination, the amplitude of the loops in the $QU$-plane decreases as the dilaton parameter $b_0$ increases, indicating a reduction in the observed polarization degree. Physically, this behavior stems from the increased gravitational redshift and light deflection in spacetimes with larger dilaton charge, which act to depolarize the signal. Conversely, for any fixed value of $b_0$, increasing the inclination enhances the observed polarization amplitude, as beaming effects amplify the radiation emitted toward the observer. This enhancement is notably stronger for the toroidal field than for the vertical one, consistent with the different alignment of the magnetic field lines relative to the observer’s line of sight. A comprehensive analysis separating the contributions of special relativity and spacetime curvature to the $QU$ diagrams can be found in~\cite{Vincent:2023sbw}.

Overall, these results show that the topology of the $QU$-loops and their evolution with inclination are very sensitive to both the magnetic field geometry and the underlying spacetime structure. 
Note also that our analytical formulas are not restricted to circular orbits. They can also be applied to study the trajectories of hot spots accreting onto the black hole (for recent studies in this context, see Refs.~\cite{Chen:2024jkm,Wang:2025btn,Xie:2025skg}).

\section{Final Remarks} \label{sec:final_remarks}
In this work, we have extended the analytical formulas of Ref.~\cite{Claros:2024atw} to spherically symmetric spacetimes in which the metric components $g_{tt}$ and $g_{rr}$ are arbitrary functions of the radial coordinate. Following the methodology of Ref.~\cite{Claros:2024atw}, the analysis reduces to evaluating the integrals defined in Eq.~\eqref{eq:int-main}. Notably, the integrals $\mathcal{I}_n$ can be computed analytically for a nontrivial family of spherically symmetric spacetimes. In cases where this is not possible, they reduce to integrals that depend only on the physical parameters of the metric and the radial position of the emission point, in contrast to the exact integral, which also depends on the emission angle.

Specifically, we have derived fully analytical expressions for two widely used parametric metric families-the Johannsen–Psaltis and Rezzolla–Zhidenko metrics-allowing a direct mapping between emission and observation points for asymptotic observers. These analytical expressions make it possible to perform fast and accurate ray-tracing computations across different gravity models. Furthermore, we have obtained compact analytical formulas for the Einstein–Maxwell–dilaton–axion spacetime and applied them to study synchrotron polarization through synthetic $QU$-loop diagrams.

The comparison with numerical integrations shows that the simplified expression \eqref{eq:our_aprox_belo} remains accurate up to observer inclinations of $\theta_o \lesssim 60^{\circ}$, whereas the refined approximation \eqref{eq:our_aprox_full} attains high accuracy for arbitrary inclinations up to angles of about $85^\circ$, including the strong-field regime close to the event horizon.
 This provides a fast and computationally inexpensive alternative to full numerical ray tracing, while retaining excellent accuracy for direct light trajectories.

Beyond the specific cases analyzed here, the generalization presented offers a flexible framework applicable to a wide range of astrophysical problems such as: thick accretion disk models, polarization pattern studies, pulsar light curves, astrophysical studies of exotic objects, among others. On the other hand, the techniques developed here can, in principle, be used to study light rays that do not necessarily follow null geodesics of the physical spacetime. For instance, in certain nonlinear electrodynamics theories, rays in the geometric--optics limit do not follow null geodesics of the background metric but instead propagate along null geodesics of an associated effective metric. Similarly, as we shown in~\cite{BG_2023}, closely related formulas can be used to treat light propagation in dispersive media, such as plasma environments.
Although the analytical formulas developed here are restricted to direct rays, recent analytical studies of higher-order images produced by equatorial emission rings provide a complementary pathway to probe strong-field lensing. In Schwarzschild spacetime, Bisnovatyi-Kogan and Tsupko analyzed the case with the observer on the disk’s symmetry axis, and Tsupko extended the treatment to arbitrary disk inclinations~\cite{Bisnovatyi-Kogan:2022ujt,Tsupko:2022kwi}. More recently, these ideas have been generalized to broader classes of spacetimes~\cite{Aratore:2024bro}. Incorporating such higher-order images within our semi-analytical framework is a natural direction for future work and would further broaden the diagnostic power of imaging and polarimetric observables in the strong-field regime.
 
Finally, as part of ongoing work, we are extending the present framework to stationary and axisymmetric spacetimes in order to carry out an analytical study of synchrotron polarization on rotating black holes with moderate spin. This effort builds upon the methodology introduced in Ref.~\cite{Loktev:2023cty} and includes a direct comparison with the \textsc{Skylight} ray-tracing code~\cite{Pelle:2022phf}, which is capable of handling both analytical and numerically tabulated metrics.

\subsubsection*{Acknowledgments}

We acknowledge financial support from CONICET, SeCyT-UNC. J.Claros is supported by a doctoral fellowship from CONICET. 
    


\end{document}